\documentclass[aps,prb,reprint,superscriptaddress,longbibliography]{revtex4-2}

\usepackage{amsmath,amssymb,amsfonts,bm,mathtools,booktabs}
\usepackage{graphicx}
\usepackage{float}
\usepackage[colorlinks=true,allcolors=blue]{hyperref}

\newcommand{\ii}{\mathrm{i}}

\newcommand{\dd}{\mathrm{d}}
\newcommand{\one}{\mathbb{I}}
\newcommand{\Om}{\Omega}
\newcommand{\calB}{\mathcal{B}}
\newcommand{\bmr}{\bm r}
\newcommand{\bmp}{\bm p}
\newcommand{\bmq}{\bm q}
\newcommand{\bmd}{\bm d}
\newcommand{\bms}{\bm \sigma}

\begin{document}

\title{Local phase-space Berry curvature and Hall transport in textured twisted bilayer graphene}
\author{Tohid Farajollahpour}
\email{tohid.farajollahpour@ntnu.no}
\affiliation{Department of Physics, Norwegian University of Science and Technology (NTNU), NO-7491 Trondheim, Norway}
\affiliation{Department of Physics, Brock University, St. Catharines, Ontario L2S 3A1, Canada}
\date{\today} 

\begin{abstract}
Slow twist-angle and heterostrain textures in twisted bilayer graphene provide a natural route to phase-space Berry geometry. We show that a purely geometric tetrad/shift sector does not generate mixed Berry curvature once the spin connection is treated consistently. By contrast, a projected textured mini-Dirac cone in twisted bilayer graphene acquires genuine mixed phase-space curvature. We analyze a controlled local Hall-bar limit with a one-dimensional texture and mirror $M_x$. In that limit the pseudogauge gradient renormalizes only the longitudinal conductivity along the textured direction, while a dc nonlinear Hall response appears only when an additional valley-odd tilt generates a Berry-curvature dipole. Both geometric responses peak at the same local filling, $\mu_{\rm loc}=\sqrt{2}\,m$, providing a gate-tunable fingerprint of their common origin. All local-cone parameters and their texture susceptibilities are extracted from a heterostrained Bistritzer-MacDonald model at $\theta=1.3^\circ$. The tilt invoked in the transport estimates corresponds to heterostrain of only $0.03\%-0.2\%$, below values routinely imaged in devices. The resulting theory provides a local analytic framework for textured moir\'e Dirac materials and cleanly separates geometric, texture-induced, and transport-level ingredients relevant to realistic TBG.
\end{abstract}

\maketitle

{\color{blue}\textit{Introduction.}}--- 
Twisted bilayer graphene (TBG) is a paradigmatic moir\'e Dirac material whose low-energy bands are highly sensitive to twist angle, lattice relaxation, heterostrain, and substrate alignment \cite{LopesDosSantos2007,SuarezMorell2010,BistritzerMacDonald2011,NamKoshino2017,CarrExact2019,Balents2019,Cao2018Insulator,Cao2018SC,Kerelsky2019,Xie2019,Choi2019,Huder2018,Mesple2021,Kazmierczak2021,Sharpe2019,Serlin2020}. More broadly, moir\'e materials provide a platform for strongly correlated, topological, and quantum-geometric phenomena \cite{Balents2020Review,Kennes2021,Andrei2021Marvels,Adak2024,AlEzzi2025}. Local reconstruction and twist-angle disorder are therefore central rather than peripheral. Microscopy, spectroscopy, and transport imaging now show that realistic TBG devices host slow spatial textures and local moir\'e inhomogeneity \cite{Yoo2019,Wilson2020Disorder,Uri2020,Kazmierczak2021,Gadelha2021,Potocnik2023,Hu2024TwistMap,QTM2023,Yu2024TBGTextures,Ciepielewski2025,Birkbeck2025,KangVafek2025}. Realistic samples should accordingly be viewed as slowly textured media in which miniband parameters vary across the device.

This immediately raises a transport question. Berry-phase semiclassics shows that slowly varying perturbations are governed by a phase-space symplectic structure rather than by momentum-space curvature alone \cite{SundaramNiu1999,XiaoNiu2010,GaoYangNiu2014}. Hall measurements are likewise established probes of Berry curvature, Berry-curvature dipoles or quadrupoles, and quantum geometry in moir\'e systems \cite{SodemannFu2015,Farajollahpour2025,Ma2019NLH,Kang2019NLH,Ma2021Review,Du2021Review,Duan2022NLH,Zhang2022NLH,Huang2023INH,Chen2024Crossed}. Yet most analyses of Hall-type responses in TBG begin from a spatially uniform miniband whose point-group symmetry is broken globally. In a textured sample, by contrast, a wave packet explores a band structure that depends on both momentum and position, so the natural geometric object is the full phase-space Berry tensor, including mixed components $\Omega_{r_i p_j}$.

It is essential, however, to separate two sources of position dependence. One is the purely geometric tetrad/shift structure of a covariant Dirac Hamiltonian. The other is the genuine projected dependence of local miniband parameters on twist angle and heterostrain. Our central result is that these are not equivalent: once the spin connection is treated consistently, the pure geometric sector cancels out of the mixed Berry curvature, whereas the projected texture dependence of the local TBG cone produces a genuine phase-space response.

We therefore organize the analysis in three steps. First, at the covariant level we prove that a pure tetrad/shift sector by itself generates no mixed Berry curvature in an isolated Dirac band. Second, at the projected-cone level we derive the local TBG miniband curvature generated by slow twist-angle and heterostrain textures. Third, at the transport level we restrict to a controlled mirror-preserving Hall-bar limit and solve the Boltzmann problem analytically only within that local setting. Finally, every local-cone coefficient and texture susceptibility used below is extracted from a heterostrained Bistritzer-MacDonald (BM) model~\cite{Supp} (see also Refs.~\cite{Jung2015,Bi2019,Vozmediano2010} therein). This yields a transport decomposition of the in-plane current density $j_i$, with $i\in\{x,y\}$, into a linear Drude part, a linear mixed-curvature correction, and a second-order Berry-curvature-dipole contribution,
\begin{equation}
 j_i=j_i^{\mathrm D}+j_i^{\mathrm{mix}}+j_i^{\mathrm{BCD}}+O(E^3),
 \label{eq:prl_current_decomposition}
\end{equation}
where $j_i^{\mathrm D}$ is the anisotropic Drude response of the local cone, $j_i^{\mathrm{mix}}$ is the linear correction induced by mixed phase-space curvature, and $j_i^{\mathrm{BCD}}$ is the nonlinear Hall current generated by the Berry-curvature dipole (BCD). The key point is that the covariant cancellation is general, the projected-cone Berry geometry is local, and the transport formulas below apply only to the specialized Hall-bar patch introduced later.

{\color{blue}\textit{Covariant origin of the mixed curvature.}}--- We begin with the underlying $4\times4$ covariant Dirac formulation.
The point is not that TBG is literally a relativistic field theory, but that a covariant representation cleanly organizes what belongs to pure geometry and what must be inserted as a physical projected correction.
Parameterized by a spatial shift field $N^i(\bmr)$ and a tetrad $e_{\hat a}{}^{\,i}(\bmr)$, the static Hamiltonian is~\cite{Supp}
\begin{equation}
\hat H_{\mathrm{min}}=\frac12\{N^i(\bmr),\Pi_i\}\one_4+\frac12\{\alpha^{\hat a}e_{\hat a}{}^{\,i}(\bmr),\Pi_i\}+m(\bmr)\beta,
\label{eq:prl_minimal_4x4_ham}
\end{equation}
where $i\in\{x,y\}$ is the spatial coordinate index, $\hat a\in\{\hat x,\hat y\}$ is the spatial local-frame index, $\Pi_i$ is the Hermitian kinetic momentum, and $m(\bmr)$ is the local Dirac mass. The operators $\alpha^{\hat a}$ and $\beta$ are the standard Dirac matrices, $\one_4$ is the $4\times4$ identity matrix.
For the purely geometric, matrix-valued tetrad sector, the corresponding $d$-vector is $d_{\hat A}=e_{\hat A}{}^{\mu}p_{\mu}$, where $e_{\hat A}{}^{\mu}$ is the full local spacetime tetrad and $p_{\mu}$ the covariant momentum. The relevant covariant real-space derivative is then
\begin{equation}
\mathcal D_i d_{\hat A}=\bigl(\nabla_i e_{\hat A}{}^{\mu}\bigr)p_{\mu}=0,
\label{eq:prl_tetrad_postulate}
\end{equation}
where the last equality is simply the tetrad postulate, including the Christoffel and spin-connection terms~\cite{Birrell_1982}. Here ``pure geometry'' means position dependence carried only by the tetrad and shift fields of the covariant Dirac operator, with no additional texture dependence in projected band parameters. 
Since the mixed curvature of an isolated Dirac band is built from $\mathcal D_i\bmd\times \partial_{p_j}\bmd$, Eq.~\eqref{eq:prl_tetrad_postulate} implies that this pure geometric tetrad/shift sector contributes nothing to $\Omega_{r_i p_j}$. Nonzero mixed curvature must therefore come from additional position dependence in the projected band data. This statement is especially important for moir\'e materials, where slowly varying twist-angle and heterostrain fields place transport in a local-moir\'e regime and make the effective miniband data explicitly position dependent. In that setting, one should not over-interpret a purely kinematic tetrad/shift geometry as an independent source of transport; instead, any nonzero mixed phase-space response must be tied to the texture dependence of the projected miniband parameters themselves \cite{Balents2019,Yoo2019,Kazmierczak2021,SundaramNiu1999,XiaoNiu2010,GaoYangNiu2014}.

{\color{blue}\textit{Local moir\'e reduction.}}---In textured TBG the natural slow fields are the local twist angle $\theta(\bmr)$ and the layer-relative heterostrain tensor $u^-_{ij}(\bmr)$. The appropriate regime is the local-moir\'e limit,
\begin{equation}
L_{\mathrm{tex}}\gg L_{\mathrm M},
\label{eq:prl_local_moire_regime}
\end{equation}
where $L_{\mathrm{tex}}$ is the texture scale and $L_{\mathrm M}$ the moir\'e period. Locally one may then assign a BM-like moir\'e band structure to each point in the sample \cite{BistritzerMacDonald2011,NamKoshino2017,Balents2019,CarrExact2019,Yoo2019}. We collect the slow texture variables into $\Phi_\alpha$ consisting of the local twist-angle deviation $\delta\theta$ together with the three independent invariants of the layer-relative strain tensor $u^-_{ij}$ as~\cite{Kazmierczak2021,Supp}
\begin{equation}
\begin{aligned}
\Phi_\alpha&=(\delta\theta,u_0,u_1,u_2),\\
u_0&=\tfrac12(u^-_{xx}+u^-_{yy}),\qquad
u_1=\tfrac12(u^-_{xx}-u^-_{yy}),\qquad
u_2=u^-_{xy}.
\end{aligned}
\label{eq:prl_texture_fields}
\end{equation}
where $\delta\theta=\theta-\theta_{\rm ref}$ measures the local deviation from a reference twist angle and the variables $u_0,u_1,u_2$ denote the three independent symmetry-adapted
components of the layer-relative heterostrain tensor $u^-_{ij}$. In this decomposition, $u_0$ is the isotropic dilation, while
$u_1$ and $u_2$ are the two independent traceless shear components.

Near an isolated valley-resolved mini-Dirac point, the projected two-band Hamiltonian $H_{\nu}$ can be written as
\begin{align}
&H_{\nu}(\bmr,\bmp)=\varepsilon_0\one_2+\bm w\!\cdot\!\bmq\,\one_2+\nu v_x q_x\sigma_x+v_y q_y\sigma_y+m\sigma_z,
\nonumber\\ &q_i=p_i-A_i^{(\nu)}(\bmr),
\label{eq:prl_local_cone}
\end{align}
where $\varepsilon_0$ is the scalar energy shift, $\bm w$ is the tilt vector, $v_x$ and $v_y$ are anisotropic Dirac velocities, $m$ is the local Dirac mass, and $\bm q$ is the momentum measured from the valley-dependent mini-Dirac point. In this projected basis $\one_2$ is the $2\times2$ identity matrix, $\sigma_{x,y,z}$ are Pauli matrices, and $\nu=\pm$ labels the two valleys. 
The smooth coefficients $\varepsilon_0$, $\bm w$, $v_x$, $v_y$, $m$, and $A_i^{(\nu)}$ are functions of the local texture fields. To parameterize how any generic local-cone coefficient $X\in\{\varepsilon_0,w_i,v_a,m,A_i^{(\nu)}\}$ responds to the texture, we write
\begin{equation}
\begin{aligned}
X(\bmr)&=X_{\rm ref}+\sum_{\alpha}X_{\alpha}\Phi_\alpha(\bmr)+\cdots,\\
\partial_i X&=\sum_{\alpha}X_\alpha\,\partial_i\Phi_\alpha+\cdots,
\end{aligned}
\label{eq:prl_texture_expansion}
\end{equation}
for any coefficient $X$ in Eq.~\eqref{eq:prl_local_cone}. The scalar shift $\varepsilon_0(\bmr)$ does not enter the Berry curvatures directly, but it shifts the local band edge and therefore the relation between the global and local electrochemical potentials that appears in transport. In the Hall-bar reduction below this scalar sector also contains the valley-even combination $-w_xA_x$ coming from $\nu w_x q_x=\nu w_x p_x-w_xA_x$. The strong sensitivity of these local cone parameters to heterostrain, relaxation, and twist-angle variation is by now well studied both theoretically and experimentally \cite{Huder2018,Mesple2021,Kazmierczak2021,Yu2024TBGTextures,KangVafek2025}. The coefficients $X_\alpha$ are not treated as free parameters. The Supplemental Material \cite{Supp} extracts all of them, including the valley-odd tilt response, from a heterostrained BM model at the reference angle, and the results are collected in its Table~S2. The transport theory below is nevertheless formulated purely at the level of the local cone, Eq.~\eqref{eq:prl_local_cone}, so it remains a controlled local-cone analysis into which refined microscopic inputs (lattice relaxation, interactions) can be inserted without structural change.


{\color{blue}\textit{Berry curvature of the local miniband.}}---For the local cone in Eq.~\eqref{eq:prl_local_cone}, the Berry curvature is given by~\cite{Supp}
\begin{equation}
\Omega_{MN}^{(\lambda,\nu)}=-\frac{\lambda}{2\varepsilon^3}\,\bmd\cdot\bigl(\partial_M\bmd\times\partial_N\bmd\bigr), 
\label{eq:pr_projector_formula}
\end{equation}

where $M,N\in\{r_x,r_y,p_x,p_y\}$ label phase-space coordinates, $\lambda=+$ ($-$) denotes the conduction (valence) band, $\bmd=(\nu v_x q_x,\, v_y q_y,\, m)$, and $\varepsilon=\sqrt{v_x^2q_x^2+v_y^2q_y^2+m^2}$. 
The momentum-space curvature follows immediately,
\begin{equation}
\Omega_{p_xp_y}^{(\lambda,\nu)}=-\frac{\lambda\nu v_xv_y m}{2\varepsilon^3}.
\label{eq:prl_pp_curvature}
\end{equation}
The mixed curvature is obtained by differentiating the local cone parameters with respect to position~\cite{Supp}, giving 
\begin{align}
\Omega_{r_i p_x}^{(\lambda,\nu)}&=-\frac{\lambda\nu v_xv_y}{2\varepsilon^3}\Bigl[q_y\,\Xi_i^{(y)}+m\,\partial_iA_y^{(\nu)}\Bigr],\notag\\
\Omega_{r_i p_y}^{(\lambda,\nu)}&=\phantom{-}\frac{\lambda\nu v_xv_y}{2\varepsilon^3}\Bigl[q_x\,\Xi_i^{(x)}+m\,\partial_iA_x^{(\nu)}\Bigr].
\label{eq:prl_mixed_curvature_general}
\end{align}
where 
$\Xi_i^{(a)}\equiv \partial_i m-m\,\partial_i\ln v_a$. There are two distinct ways for a texture to enter the mixed curvature. The first is through gradients of the local gap and the local anisotropic velocities, encoded by $\Xi_i^{(a)}$. The second is through gradients of the mini-Dirac-point shift $A_i^{(\nu)}$, which act as moir\'e pseudogauge fields. The effective tilt $\bm w$ does not enter the curvature directly. Instead, it enters transport indirectly through the shape of the Fermi surface.

\begin{figure}[t]
 \centering
    \includegraphics[width=0.98\linewidth]{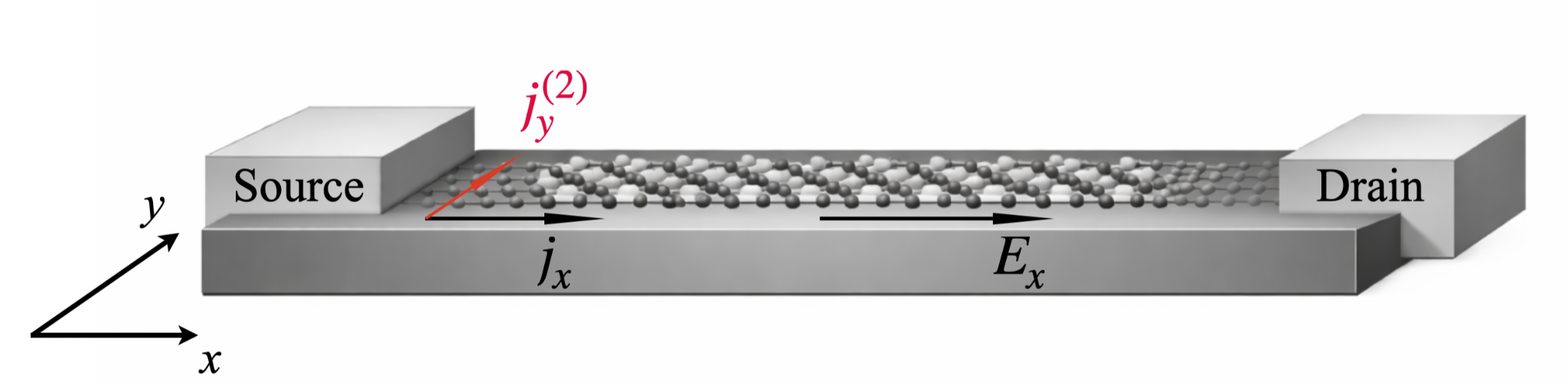}
    \caption{Schematic of the controlled local Hall-bar patch. The slow texture varies along $y$ across the bar, while the drive $E_x$ is applied along $x$. The linear response is longitudinal, $j_x^{(1)}$, and the nonlinear Hall signal is the second-order transverse current $j_y^{(2)}$. A drive $E_y$ instead probes the textured longitudinal conductivity.}
    \label{fig1}
\end{figure}

{\color{blue}\textit{TBG Hall-bar scope and symmetry filter.}}---From this point onward the analytic transport formulas refer to a controlled local Hall-bar limit (see Fig.~\ref{fig1}). We choose the transport axes so that the slow texture is one-dimensional, $\partial_y\Phi_\alpha\neq 0$ and $\partial_x\Phi_\alpha=0$, and the sample preserves the mirror symmetry 
\begin{equation}
M_x:(x,y)\mapsto(-x,y).
\label{eq:prl_hallbar_mirror_intro}
\end{equation}
The minimal local cone consistent with this geometry keeps a scalar shift $\varepsilon_0$, a valley-odd tilt $w_x$ along the bar, and a valley-dependent pseudogauge shift $A_x$ of the Dirac point, with $A_y=0$. We further restrict to a local patch in which $
\partial_y\mu_{\rm loc}=\partial_y w_x=\partial_y m=\partial_y v_x=\partial_y v_y=0
$,
so the valley-even scalar sector is organized as
\begin{equation}
\varepsilon_{\rm sc}(y)\equiv \varepsilon_0(y)-w_x(y)A_x(y),
\qquad
\mu_{\rm loc}(y)=\mu_{\rm glob}-\varepsilon_{\rm sc}(y),
\label{eq:Shift}
\end{equation}
and the only retained gradient is the pseudogauge field
\begin{equation}
 \mathcal B_s(y)\equiv \partial_yA_x(y).
\label{eq:prl_pseudomagnetic}
\end{equation}
Physically, this limit isolates the dominant kinematic effect of the texture, and the microscopic extraction in the Supplemental Material \cite{Supp} makes the statement quantitative. For shear-strain textures ($u_1$, $u_2$) the competing gap- and velocity-gradient channel $\Xi_i^{(a)}$ is forbidden at linear order by the $C_{3z}$ symmetry of the reference structure (Table~S2 \cite{Supp}), so the pseudogauge gradient is parametrically dominant. For twist-angle textures the two channels coexist. The ratio $k_F\,\partial_\theta m/(m\,|\partial_\theta \bm A|)$ grows from $\approx0.4$ at $\mu_{\rm loc}=1.1\,m$ to $\approx0.9$ at $\mu_{\rm loc}=\sqrt{2}\,m$, so the pseudogauge term dominates near the band edge, and Eq.~\eqref{eq:prl_mixed_curvature_general} retains the $\Xi$ terms so that they can be reinstated without structural change. The local mini-Dirac point shift $A_x$ is directly tied to the geometry of the moir\'e Brillouin zone (tracking, for instance, the twist-angle gradient $\partial_y\theta$), and its resulting pseudomagnetic field $\mathcal B_s$ then constitutes the leading-order phase-space perturbation. Suppressing the remaining gradients of $m$ and $v_a$ cleanly disentangles the transport signatures of the pseudogauge field from those generated by $\Xi_i^{(a)}$. This reduction is consistent with recent symmetry analyses of twisted moir\'e materials \cite{AlEzzi2025}. In this geometry, Fig.~\ref{fig1}, $E=(E_x,0)$ probes the ordinary longitudinal current $j_x^{(1)}$ together with any second-order transverse response $j_y^{(2)}$, whereas $E=(0,E_y)$ isolates the textured longitudinal channel $j_y^{(1)}$~\cite{Supp}.

Within this scope the only mixed curvature needed below reduces to
\begin{equation}
 \Omega_{yp_y}^{(+,\nu)}=\frac{v_xv_y m}{2\varepsilon^3}\,\mathcal B_s, \qquad \Omega_{yp_x}^{(+,\nu)}=0.
 \label{eq:prl_specific_mixed_curvature}
\end{equation}
The mirror transformation acts on polar vectors as $j_x\to-j_x$, $j_y\to j_y$, $E_x\to -E_x$, and $E_y\to E_y$. Consequently the linear Hall tensor is forbidden,
 \begin{equation}
  \sigma_{xy}=\sigma_{yx}=0, \label{eq:prl_linear_filter}
\end{equation}
while at second order symmetry allows only the components of the nonlinear conductivity tensor $\chi_{ilm}$ defined by $j_i^{(2)}=\chi_{ilm}E_lE_m$,
\begin{equation}
 \chi_{yxx},\qquad \chi_{xxy}=\chi_{xyx},\qquad \chi_{yyy}.
 \label{eq:prl_nonlinear_filter}
\end{equation}
Here $\chi_{yxx}$ is the nonlinear Hall coefficient relevant for a pure longitudinal drive $E=(E_x,0)$, $\chi_{xxy}=\chi_{xyx}$ requires both field components to be present simultaneously, and $\chi_{yyy}$ is symmetry-allowed at this stage but will be shown below to vanish in the minimal local Hall-bar cone.

{\color{blue}\textit{Boltzmann transport and current decomposition.}}---The transport calculation is performed in a local-frame relaxation-time approximation for a conduction-band Fermi energy $\mu_{\rm loc}>|m|$. Before the local specialization, the steady kinetic equation has the standard spatial form $\dot r_i\partial_{r_i}f_\nu+\dot p_i\partial_{p_i}f_\nu=-(f_\nu-f_\nu^{\rm loc})/\tau$, and $\dot p_y$ contains both electric forcing and a scalar-force contribution $-\partial_yE_{+,\nu}(\bmr,\bmp)$. In the Hall-bar patch defined above, the local electrochemical equilibrium $f_\nu^{\rm loc}$ is taken to absorb the valley-even scalar sector $\varepsilon_{\rm sc}$ defined in Eq.~\eqref{eq:Shift}, while the remaining gradients of $w_x$, $m$, and $v_a$ are neglected. The closed formulas below therefore retain only the pseudogauge gradient $\partial_yA_x$ inside the Berry-geometric sector. Under that controlled approximation the phase-space dynamics naturally separate the current into the three pieces already mentioned in Eq.~\eqref{eq:prl_current_decomposition}. The Drude term comes from the group velocity and the nonequilibrium correction of the distribution function in the local anisotropic band. The mixed-curvature term comes from the modification of the semiclassical velocity and the invariant phase-space measure by $\Omega_{r_ip_j}$, as detailed in the Supplemental Material. The BCD term is the second-order current generated by the dipole of the ordinary momentum-space curvature $\Omega_{p_xp_y}$. 


Writing $\mu_{\rm loc}>|m|$ for the local conduction-band chemical potential, $\tau$ for the relaxation time, and $g_s$ and $g_v$ for the spin and valley degeneracies, the first-order solution of the Boltzmann equation gives the familiar anisotropic Drude response plus a linear correction from the mixed curvature. After integrating over the Fermi contour~\cite{Supp}, we obtain
\begin{align}
\sigma_{xx}&=\frac{g_sg_v e^2\tau}{4\pi}\,\frac{v_x}{v_y}\,\frac{\mu_{\rm loc}^2-m^2}{\mu_{\rm loc}},\notag\\
\sigma_{yy}&=\frac{g_sg_v e^2\tau}{4\pi}\,\frac{v_y}{v_x}\,\frac{\mu_{\rm loc}^2-m^2}{\mu_{\rm loc}}\left(1+\frac{v_xv_y m\,\mathcal B_s}{2\mu_{\rm loc}^3}\right),
\label{eq:prl_linear_sigma}
\end{align}

In this $M_x$-preserving limit the linear Hall tensor vanishes, while mixed curvature renormalizes only the longitudinal conductivity along the textured direction. The correction scales as $v_xv_y m\mathcal B_s/(2\mu_{\rm loc}^3)$, so its sign can be reversed either by changing the gap or by reversing the texture gradient that determines $\mathcal B_s$.

We denote the Drude conductivities of the local cone by $\sigma_{aa}^{\rm D}$, the mixed-curvature correction by $\delta\sigma_{yy}^{\rm mix}$, and the second-order conductivity generated by the Berry-curvature dipole by $\chi_{ilm}^{\rm BCD}$. For the pure-$E_x$ and pure-$E_y$ drives emphasized below,
\begin{equation}
\begin{aligned}
j_x&=\sigma_{xx}^{\rm D}E_x+O(E_xE_y,E^3),\\
j_y&=\bigl(\sigma_{yy}^{\rm D}+\delta\sigma_{yy}^{\rm mix}\bigr)E_y+\chi_{yxx}^{\rm BCD}E_x^2+O(E_xE_y,E^3),
\end{aligned}
\label{eq:prl_explicit_current_decomposition}
\end{equation}
with $\sigma_{xx}^{\rm D} = \sigma_{xx}$, $\sigma_{yy}=\sigma_{yy}^{\rm D}+\delta\sigma_{yy}^{\rm mix}$, and
\begin{equation}
\sigma_{yy}^{\rm D}=\frac{g_sg_v e^2\tau}{4\pi}\,\frac{v_y}{v_x}\,\frac{\mu_{\rm loc}^2-m^2}{\mu_{\rm loc}},
\qquad
\delta\sigma_{yy}^{\rm mix}=\sigma_{yy}^{\rm D}\,\frac{v_xv_y m\mathcal B_s}{2\mu_{\rm loc}^3}.
\label{eq:prl_mixed_sigma_term}
\end{equation}
Equation~\eqref{eq:prl_explicit_current_decomposition} keeps only the coefficients relevant to the pure-$E_x$ and pure-$E_y$ drives; the crossed coefficients $\chi_{xxy}=\chi_{xyx}=-\chi_{yxx}/2$ are given explicitly in Eq.~\eqref{eq:prl_nlh}, while $\chi_{yyy}=0$ in the minimal local Hall-bar cone.

\begin{figure}[!t]
 \centering
    \includegraphics[width=0.99\linewidth]{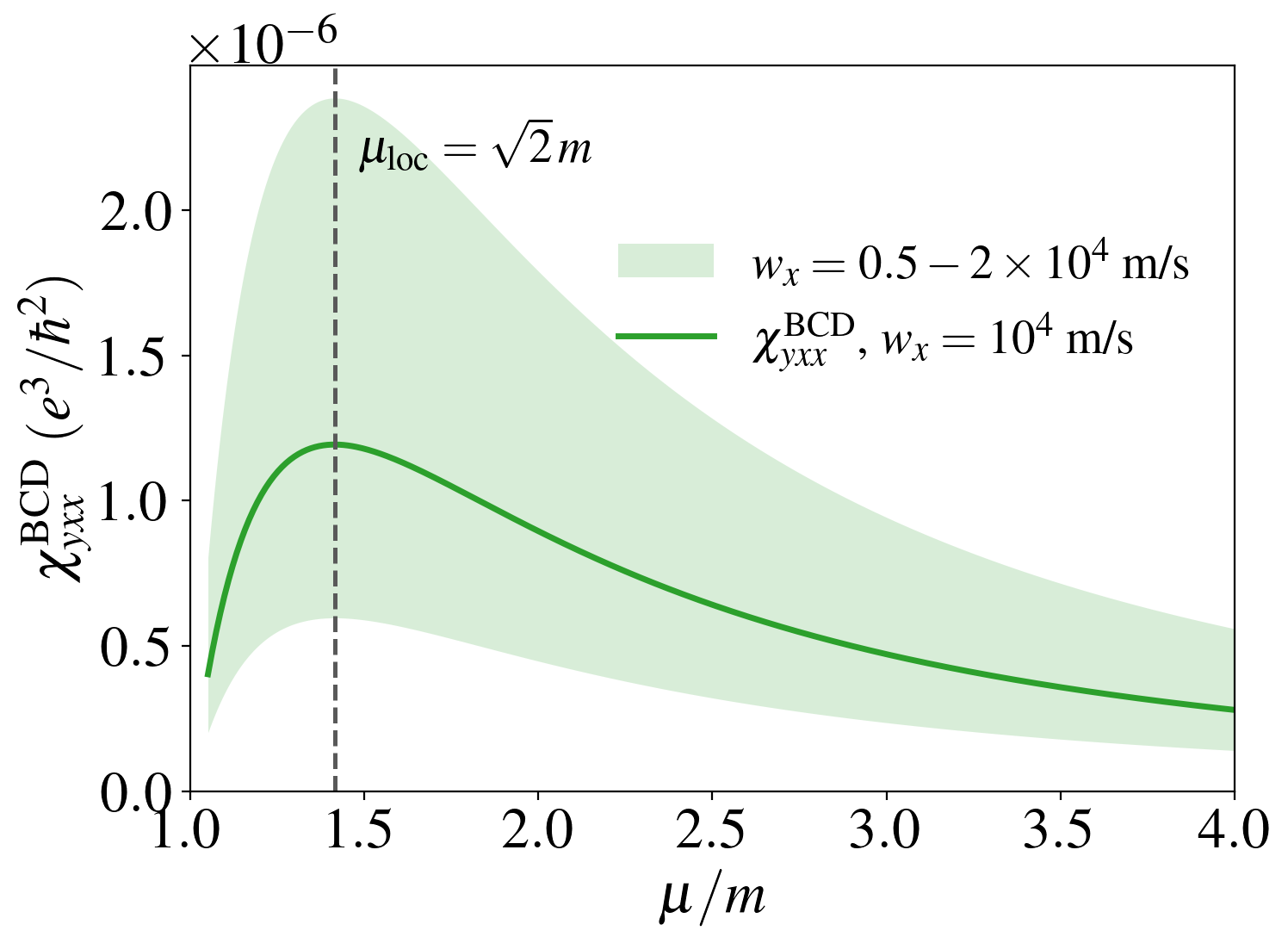}
    \caption{The dc nonlinear Hall coefficient $\chi_{yxx}^{\mathrm{BCD}}$ generated by the valley-odd tilt, for $m = 10$~meV, $\tau = 10$~ps, and $g_s=g_v=2$. The solid line uses $w_x = 10^4$~m/s, which the Bistritzer-MacDonald extraction in the Supplemental Material \cite{Supp} identifies with a uniaxial heterostrain $\varepsilon\simeq0.06\%-0.14\%$, depending on the strain orientation. The shaded band covers the conservative range $w_x=0.5-2\times10^4$~m/s, corresponding to $\varepsilon\simeq0.03\%-0.2\%$. At a representative $\varepsilon=0.3\%$ the extracted tilt reaches $4-7\times10^4$~m/s [Fig.~S2(b) of the Supplemental Material], so the plotted response is a conservative lower end. The dashed line marks $\mu_{\rm loc}=\sqrt{2}\,m$, where $\chi_{yxx}$ peaks together with $\delta\sigma_{yy}^{\rm mix}$ (Fig.~\ref{fig3}).}
           \label{fig2}
\end{figure}

\begin{figure}[!t]
 \centering
    \includegraphics[width=0.99\linewidth]{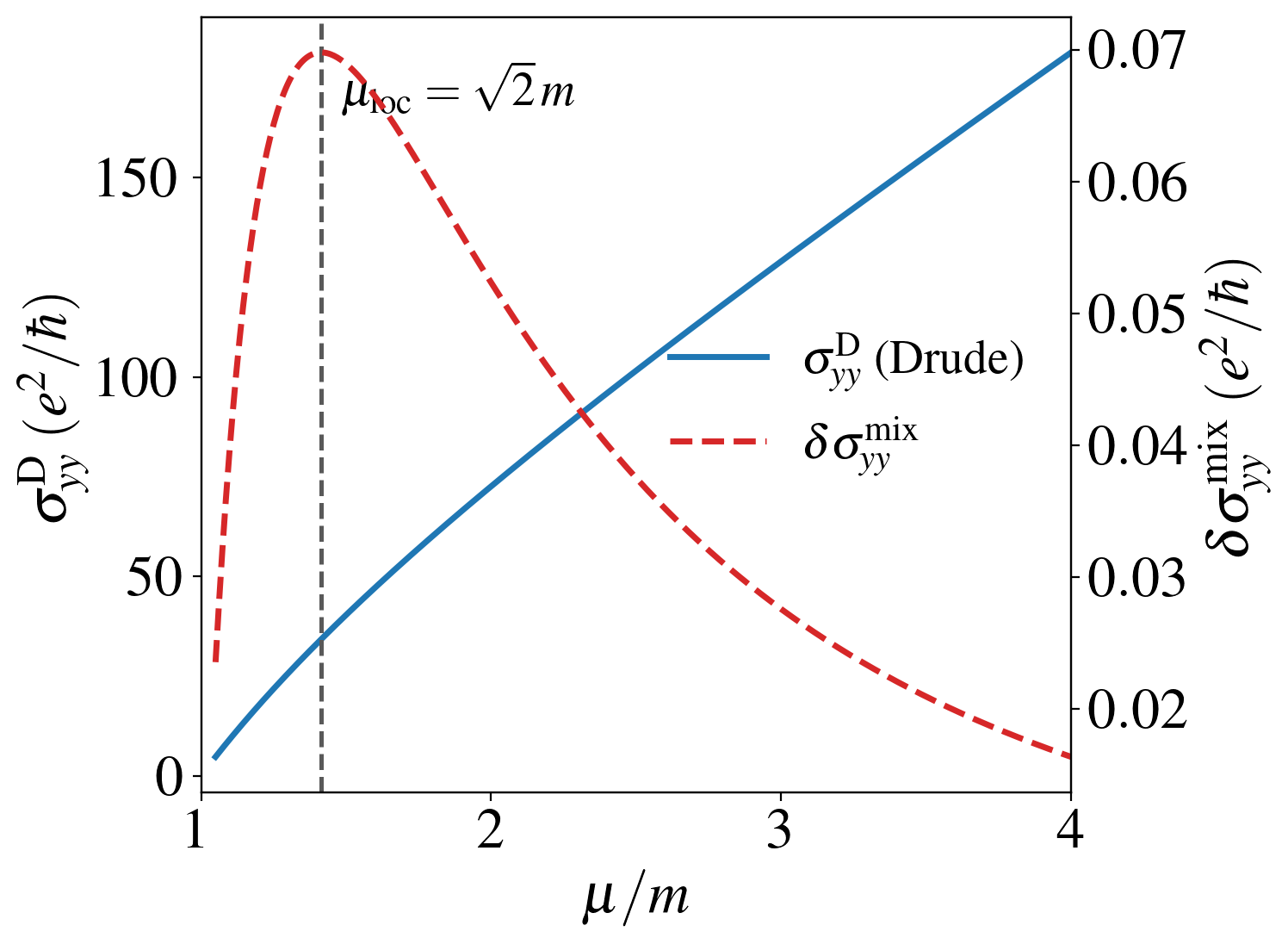}
    \caption{The linear longitudinal conductivity components along the textured direction, for the BM-extracted reference velocities $v_x = v_y = 1.63 \times 10^5$~m/s (Supplemental Material \cite{Supp}), $m = 10$~meV, $\tau = 10$~ps, and $\mathcal{B}_s\equiv\partial_yA_x = 10^{14}$~m$^{-2}$, with $g_s=g_v=2$. By Table~S2 of the Supplemental Material \cite{Supp}, this $\mathcal B_s$ corresponds, e.g., to a twist gradient $|\partial_y\theta|\simeq0.7^\circ/\mu$m or to a shear-strain gradient $\partial_yu_1\simeq1.3\times10^{-3}/\mu$m, both within imaged texture amplitudes~\cite{Uri2020,Kazmierczak2021}. The dashed line marks $\mu_{\rm loc}=\sqrt{2}\,m$, the common maximum of $\delta\sigma_{yy}^{\rm mix}$ and $\chi_{yxx}$ (Fig.~\ref{fig2}).}
       \label{fig3}
\end{figure}

{\color{blue}\textit{Nonlinear Hall response and the role of the tilt.}}---The second-order dc response is controlled by the BCD rather than directly by the mixed curvature. For a two-dimensional band, the BCD sector is expressed in terms of the dipole vector $D_a$, the second-order conductivity $\chi_{ilm}^{\rm BCD}$, the two-dimensional Levi-Civita tensor $\epsilon_{ij}$ with $\epsilon_{xy}=1$, and the local equilibrium distribution $f_0$ as
\begin{equation}
\begin{aligned}
\chi_{ilm}^{\rm BCD}&=\frac{e^3\tau}{2}\bigl(\epsilon_{il}D_m+\epsilon_{im}D_l\bigr),\\
D_a&=g_s\sum_{\nu}\int\!\frac{d^2p}{(2\pi)^2}\,\partial_{p_a}\Omega_{p_xp_y}^{(+,\nu)}f_0.
\end{aligned}
\label{eq:prl_bcd_formula}
\end{equation}
Here $g_s=2$ is the spin degeneracy, and the valley sum is kept explicit in $D_a$; after carrying it out we write the resulting factor as $g_v=2$.
In the untilted cone the integrand is parity odd and the dipole vanishes. A dc nonlinear Hall current therefore does not follow from mixed curvature alone. To obtain a nonzero dipole one needs a Fermi-surface asymmetry, which in the present TBG geometry is supplied by the valley-odd effective tilt $w_x$. Expanding the local equilibrium distribution to first order in $w_x$ and evaluating the resulting Fermi-surface integral~\cite{Supp}

\begin{equation}
D_x=-\frac{g_sg_v\,3mw_x}{8\pi}\,\frac{\mu_{\rm loc}^2-m^2}{\mu_{\rm loc}^4},
\qquad
D_y=0,
\label{eq:prl_bcd}
\end{equation}
and therefore
\begin{align}
\chi_{yxx}&=\frac{g_sg_v\,3e^3\tau mw_x}{8\pi}\,\frac{\mu_{\rm loc}^2-m^2}{\mu_{\rm loc}^4},\notag\\
\chi_{xxy}&=\chi_{xyx}=-\frac12\chi_{yxx},
\qquad
\chi_{yyy}=0.
\label{eq:prl_nlh}
\end{align}
Equation~\eqref{eq:prl_nlh} shows that the dc nonlinear Hall channel is not generated by mixed curvature alone. It appears only when a valley-odd tilt, which may contain both background and texture-induced pieces $w_x=w_x^{(0)}+\sum_\alpha w_{x,\alpha}\Phi_\alpha+\cdots$, converts momentum-space Berry curvature into a nonzero dipole~\cite{Supp}. Figure~\ref{fig2} displays the dc nonlinear Hall coefficient, $\chi_{yxx}^{\mathrm{BCD}}$. Unlike the linear transport channels, this second-order transverse response is strictly mediated by the Berry-curvature dipole and requires a valley-odd tilt to generate the requisite Fermi-surface asymmetry. Equations~\eqref{eq:prl_mixed_sigma_term} and \eqref{eq:prl_nlh} moreover share the Fermi-surface profile $(\mu_{\rm loc}^2-m^2)/\mu_{\rm loc}^4$. Both geometric responses peak at the same local filling, $\mu_{\rm loc}=\sqrt{2}\,m$, and then decay, whereas the Drude background grows monotonically (Figs.~\ref{fig2} and \ref{fig3}). The coincidence of the maxima of $\delta\sigma_{yy}^{\rm mix}$ and of the second-harmonic Hall signal at one and the same gate voltage is therefore a sharp, gate-tunable fingerprint of their common phase-space-geometric origin, distinguishing both from disorder contributions with unrelated density dependence.

\begin{table*}[t]
\caption{Symmetry-filtered transport coefficients for the local $M_x$-preserving TBG Hall-bar geometry. The Drude sector is present even without texture, the mixed sector is linear in the texture-induced pseudomagnetic field $\mathcal B_s=\partial_yA_x$, and the BCD sector requires a valley-odd effective tilt $w_x$.}
\label{tab:prl_summary}
\begin{ruledtabular}
\begin{tabular}{@{}p{0.14\textwidth}p{0.22\textwidth}p{0.38\textwidth}p{0.18\textwidth}@{}}
Coefficient & Analytic value & Physical origin & Status \\
\colrule
$\sigma_{xx}$ & $\sigma_{xx}^{\rm D}$ & anisotropic Drude response of local cone & survives \\
$\sigma_{yy}$ & $\sigma_{yy}^{\rm D}+\delta\sigma_{yy}^{\rm mix}$ & Drude baseline plus mixed phase-space curvature & survives \\
$\sigma_{xy},\sigma_{yx}$ & $0$ & forbidden by $M_x$ and two-valley time reversal & vanish \\
$\chi_{yxx}$ & $\dfrac{g_sg_v\,3e^3\tau mw_x}{8\pi}\dfrac{\mu_{\rm loc}^2-m^2}{\mu_{\rm loc}^4}$ & Berry-curvature dipole generated by valley-odd tilt & survives \\
$\chi_{xxy}=\chi_{xyx}$ & $-\chi_{yxx}/2$ & same BCD sector as $\chi_{yxx}$ & survives \\
$\chi_{yyy}$ & $0$ & symmetry-allowed but parity-odd in the minimal cone & vanishes dynamically \\
\end{tabular}
\end{ruledtabular}
\end{table*}

Table~\ref{tab:prl_summary} summarizes the hierarchy of responses. The Drude coefficients set the baseline local conductivity. The mixed phase-space curvature enters as a texture-controlled renormalization of $\sigma_{yy}$, while the BCD sector appears only in the quadratic response and only after the local miniband acquires a valley-odd tilt. The table should therefore be read as a local Hall-bar summary, not as a device-averaged transport classification for arbitrary textured TBG.

The measurement schemes suggested by these formulas are straightforward. In the $M_x$-preserving Hall-bar geometry defined above, a longitudinal drive $E=(E_x,0)$ yields a longitudinal current $j_x$ and a second-order transverse current $j_y$ according to
\begin{equation}
E=(E_x,0):\qquad j_x=\sigma_{xx}E_x,\qquad j_y=\chi_{yxx}E_x^2,
\label{eq:prl_measurement}
\end{equation}
while a drive $E=(0,E_y)$ gives $j_x=0$ and $j_y=\sigma_{yy}E_y$. Equation~\eqref{eq:prl_measurement} should be read as the short-circuit transverse response. A first-harmonic measurement probes the Drude baseline and the mixed-curvature correction in Eq.~\eqref{eq:prl_linear_sigma}. A second-harmonic transverse measurement isolates $\chi_{yxx}$. If the transverse direction is instead open circuit, one imposes $j_y=0$ and obtains the compensating second-harmonic transverse field $E_y^{2\omega}$ and the corresponding voltage drop $V_y^{2\omega}$ across probe separation $W$:
\begin{equation}
E_y^{2\omega}=-\frac{\chi_{yxx}}{\sigma_{yy}^{\rm D}+\delta\sigma_{yy}^{\rm mix}}\,E_x^2,
\qquad
V_y^{2\omega}=-W E_y^{2\omega}.
\label{eq:prl_open_circuit}
\end{equation}
The theory therefore predicts both a short-circuit current $I_{2\omega}=W\chi_{yxx}E_x^2$ and the corresponding open-circuit Hall voltage. Because $\delta\sigma_{yy}^{\rm mix}\propto m\mathcal B_s$ and $\chi_{yxx}\propto mw_x$, tuning the displacement field or hBN alignment that opens the gap, together with gate control of the local Fermi energy, provides a direct set of sign and scaling tests. Local microscopy can then correlate the transport coefficients with measured twist-angle gradients, heterostrain patterns, and moir\'e collective modes \cite{Kazmierczak2021,Yu2024TBGTextures,Birkbeck2025}.

{\color{blue}\textit{Order-of-magnitude estimates.}}---
The local mixed-curvature correction is controlled by
\begin{equation*}
\frac{\delta\sigma_{yy}^{\rm mix}}{\sigma_{yy}^{\rm D}}
=
\frac{v_x v_y\, m\, \mathcal B_s}{2\mu_{\rm loc}^3}.
\end{equation*}
In an inhomogeneous device the externally controlled chemical potential $\mu_{\rm glob}$ is shifted locally by the scalar sector, Eq.~\eqref{eq:Shift}.  
Figure~\ref{fig3} illustrates the distinct scaling behaviors of the linear conductivity components along the textured direction ($y$) as the local chemical potential $\mu_{\rm loc}$ is tuned away from the mass gap $m$. The baseline Drude conductivity, $\sigma_{yy}^{\mathrm{D}}$, exhibits the expected monotonic growth as the Fermi surface expands, scaling proportionally to $(\mu_{\rm loc}^2-m^2)/\mu_{\rm loc}$. 

In contrast, the texture-induced geometric correction, $\delta\sigma_{yy}^{\mathrm{mix}}$, is highly concentrated near the band edge. Because the mixed phase-space curvature diverges near the mini-Dirac point, the correction scales as $(\mu_{\rm loc}^2-m^2)/\mu_{\rm loc}^4$. Consequently, it reaches a sharp maximum at $\mu_{\rm loc}=\sqrt{2}\,m$ before decaying rapidly as the chemical potential increases. As indicated by the differing axis scales, this phase-space contribution constitutes a per-mille-to-sub-percent correction for representative local patches.  More favorable local patches with smaller $\mu_{\rm loc}$ and sharper texture gradients can enhance this ratio further, potentially into the percent-level regime.

High-quality $1.3^\circ$ TBG devices exhibiting intrinsic nonlinear Hall signals reach mobilities of order $1.1\times10^5\,{\rm cm^2\,V^{-1}\,s^{-1}}$ \cite{Huang2023INH}; for local cone velocities of order $10^5-2\times10^5\,{\rm m\,s^{-1}}$ and $\mu_{\rm loc}\sim20\,{\rm meV}$ this corresponds to $\tau\sim5-20\,{\rm ps}$, consistent with metallic-regime transport estimates \cite{Sharma2021Transport}. Taking the valley-odd tilt range $w_x\sim0.5-2\times10^4\,{\rm m\,s^{-1}}$, Eq.~\eqref{eq:prl_nlh} gives $|\chi_{yxx}|\sim10^{-10}-10^{-9}\,{\rm A\,m\,V^{-2}}$ for the two-dimensional sheet coefficient. In a $W\sim5\,\mu{\rm m}$ Hall bar driven at $E_x\sim10^3-2\times10^3\,{\rm V\,m^{-1}}$, this corresponds in the short-circuit limit to $I_{2\omega}=W\chi_{yxx}E_x^2\sim1-20\,{\rm nA}$, directly accessible to lock-in detection. These numbers should be viewed as illustrative local tilted-cone estimates once a valley-odd tilt is present. 
Microscopically, such a valley-odd tilt arises when heterostrain breaks the $C_{3z}$ symmetry of the moir\'e superlattice, the mechanism behind the giant strain-induced nonlinear Hall effects reported in TBG~\cite{Zhang2022NLH,Duan2022NLH}. The Bistritzer-MacDonald extraction in the Supplemental Material \cite{Supp} makes this quantitative for the present local cone. The magnitude $|w_x|$ reaches $0.5-2\times10^4\,{\rm m\,s^{-1}}$ already for uniaxial heterostrain $\varepsilon\simeq0.03\%-0.2\%$, depending on the strain orientation, while a representative $\varepsilon=0.3\%$ yields $|w_x|\simeq4-7\times10^4\,{\rm m\,s^{-1}}$ [Fig.~S2(b) \cite{Supp}]. Since residual heterostrain of $0.1\%-0.7\%$ is routinely imaged in TBG devices~\cite{Kerelsky2019,Huder2018,Mesple2021,Kazmierczak2021}, the tilt range adopted above sits at the conservative lower end of the microscopic response, and the resulting $1-20\,{\rm nA}$ short-circuit current is an experimentally realistic target. 

A useful way to organize comparison with experiment is to translate the transport coefficients back into texture susceptibilities of the local mini-Dirac cone. Here $A_{x,\alpha}$ and $w_{x,\alpha}$ are the linear susceptibilities of the local pseudogauge shift $A_x$ and valley-odd tilt $w_x$ to the texture fields $\Phi_\alpha$. In the present Hall-bar geometry,
\begin{equation}
\begin{aligned}
\mathcal B_s&=\partial_yA_x=\sum_\alpha A_{x,\alpha}\,\partial_y\Phi_\alpha+\cdots,\\
w_x&=w_x^{(0)}+\sum_\alpha w_{x,\alpha}\Phi_\alpha+\cdots,
\end{aligned}
\label{eq:prl_texture_to_transport}
\end{equation}
so the leading coefficients obey the schematic relations
\begin{equation}
\delta\sigma_{yy}^{\rm mix}\propto m\sum_\alpha A_{x,\alpha}\,\partial_y\Phi_\alpha,
\qquad
\chi_{yxx}^{\rm BCD}\propto m\,w_x.
\label{eq:prl_scaling_summary}
\end{equation}
The linear channel is therefore controlled by \emph{gradients} of the local twist-angle and heterostrain fields, whereas the nonlinear Hall channel is controlled by the \emph{local} valley-odd tilt produced by those same fields and by other symmetry-lowering perturbations. This difference is experimentally valuable. If one images a TBG device and observes strong twist-angle gradients but weak evidence for local cone tilt, then the present theory predicts a sizable correction to $\sigma_{yy}$ without a correspondingly large dc nonlinear Hall current. Conversely, a strong quadratic Hall response should be interpreted as evidence not merely for texture, but for a texture-induced or substrate-induced Fermi-surface asymmetry encoded in $w_x$.

Because $\delta\sigma_{yy}^{\rm mix}\propto m\mathcal B_s$ and $\chi_{yxx}\propto mw_x$, both signals are intrinsically local and can change sign across an inhomogeneous device. 
In a realistic 2D macroscopic sample featuring complex twist-angle networks or random domain orientations~\cite{Uri2020,Hu2024TwistMap}, a global measurement probes a weighted spatial average. Specifically, because the pseudogauge gradient $\mathcal B_s$ flips sign depending on the local texture slope, patches with opposite signs of $m(\bmr)\mathcal B_s(\bmr)$ will heavily cancel, likely washing out the linear mixed-curvature correction in macroscopic transport. The quadratic BCD response similarly averages according to the sign structure of $m(\bmr)w_x(\bmr)$. The relevant scales make this quantitative. Imaged twist-angle and heterostrain textures vary on correlation lengths $\xi\sim0.1-0.5\,\mu$m~\cite{Uri2020,Kazmierczak2021,Hu2024TwistMap}, so for a random texture the residual rms of a signed quantity such as $\mathcal B_s$ over a channel of length $L\gg\xi$ is suppressed by $\sqrt{\xi/L}$. A $5\,\mu$m bar then retains only $\sim15\%-30\%$ of the local signal, a mesoscopic $L\sim1\,\mu$m device retains $\sim50\%-70\%$, and a scanning probe with $L\lesssim\xi$ measures the full local response, while deliberately engineered monotonic gradients (bend- or tear-and-stack twist ramps) evade the suppression altogether. Because $\chi_{yxx}$ appears at the second harmonic and $\delta\sigma_{yy}^{\rm mix}$ is odd under reversing either the gap or the texture gradient, both remain separable from the smooth Drude background after partial averaging. Quantifying the full 2D network averaging requires a device-scale network or continuum calculation~\cite{Ciepielewski2025} beyond the present local theory. Therefore, to avoid signal cancellation, the most direct targets of Eqs.~\eqref{eq:prl_measurement} and \eqref{eq:prl_open_circuit} are local scanning transport probes or mesoscopic Hall bars deliberately engineered to host a largely monotonic texture gradient over their active region.

In this work we assume a locally defined TBG cone, weak enough fields for semiclassical dynamics, a local-frame treatment of the scalar sector, and a relaxation-time approximation that isolates the intrinsic geometric structure. Beyond this minimal regime, several extensions are possible. A genuinely two-dimensional texture activates the remaining mixed components $\Omega_{xp_x}$ and $\Omega_{xp_y}$ of Eq.~\eqref{eq:prl_mixed_curvature_general}, which enter the Boltzmann problem with the same structure and at the same order in gradients as $\Omega_{yp_y}$ and therefore modify the longitudinal conductivities quantitatively without changing the qualitative picture. A linear charge Hall response remains forbidden by time reversal for any static texture, independent of dimensionality. The real-space curvature $\Omega_{r_xr_y}$, which is quadratic in the texture gradients and valley odd at leading order, instead drives a valley Hall response. Relaxing $M_x$ by allowing $\partial_x\Phi_\alpha\neq0$ unfreezes the remaining second-order coefficients (for example a nonzero $\chi_{yyy}$) at $O(\partial_x\Phi_\alpha)$, with magnitudes controlled by the same susceptibilities of Table~S2 \cite{Supp}. For generic disorder with $\partial_x\Phi\sim\partial_y\Phi$ these are comparable to the coefficients computed here, while the qualitative structure, namely the common $\sqrt{2}\,m$ peak and the valley-odd tilt origin of the BCD, is unchanged. Intervalley scattering or explicit mirror breaking would similarly unfreeze the linear Hall terms that vanish in the geometry of Eq.~\eqref{eq:prl_hallbar_mirror_intro}. Extrinsic skew-scattering and side-jump mechanisms can dress the nonlinear Hall response, as emphasized in recent TBG experiments, while strong-coupling flat-band physics can renormalize the local cone parameters themselves.

Our results also clarify how the present phase-space picture connects to the broader nonlinear Hall literature. Previous TBG analyses emphasized how global symmetry lowering can generate large Berry-curvature dipoles and giant second-order responses \cite{Zhang2022NLH,Duan2022NLH,Huang2023INH,Chen2024Crossed}, in a broader context shaped by general nonlinear-Hall theory and experiment \cite{SodemannFu2015,Ma2019NLH,Kang2019NLH,Ma2021Review,Du2021Review}. The textured problem studied here is complementary. We show that slow moir\'e inhomogeneity produces a mixed phase-space curvature even when the dc nonlinear Hall effect still vanishes. The linear response is therefore already a probe of phase-space geometry, while the nonlinear Hall response requires an additional valley-odd tilt. This separation should be useful when interpreting experiments in which local twist-angle disorder, heterostrain, and substrate effects coexist and compete.

{\color{blue}\textit{Conclusions.}}---
In summary, pure geometry alone does not generate mixed curvature once the spin connection is included consistently. At the level of a local projected TBG mini-Dirac cone, twist-angle and heterostrain textures generate explicit mixed Berry curvature through the local gap, anisotropic velocities, and mini-Dirac-point shift. In the controlled $M_x$-preserving Hall-bar limit, that mixed curvature renormalizes only the longitudinal conductivity along the textured direction, while a dc nonlinear Hall response requires an additional valley-odd tilt that produces a Berry-curvature dipole. The resulting transport formulas provide a local analytic framework for textured moir\'e Dirac materials. The texture susceptibilities entering these formulas are supplied microscopically in the Supplemental Material \cite{Supp} by a heterostrained Bistritzer-MacDonald extraction, which anchors the assumed tilt and pseudogauge scales to sub-percent heterostrain and to imaged twist gradients. Extending the framework to realistic devices will require a controlled spatial kinetic treatment and device-scale averaging over measured texture networks.

{\color{blue}\textit{Acknowledgments}}---The author is grateful to R. Ganesh, S. A. Jafari, A. Qaiumzadeh and M. Salehi for helpful discussions. This work was supported by the Research Council of Norway through its Centers of Excellence funding scheme, Project No. 353919 and Project No. 361800 “QTransMag.”

\bibliography{Refs}

\clearpage
\onecolumngrid

\setcounter{equation}{0}
\setcounter{section}{0}
\setcounter{subsection}{0}
\setcounter{table}{0}
\setcounter{figure}{0}
\setcounter{secnumdepth}{3}
\renewcommand{\theequation}{S\arabic{equation}}
\renewcommand{\thesection}{S\Roman{section}}
\renewcommand{\thesubsection}{\thesection.\Alph{subsection}}
\renewcommand{\thetable}{S\arabic{table}}
\renewcommand{\thefigure}{S\arabic{figure}}
\renewcommand{\theHequation}{S\arabic{equation}}
\renewcommand{\theHsection}{S\Roman{section}}
\renewcommand{\theHsubsection}{S\Roman{section}.\Alph{subsection}}
\renewcommand{\theHtable}{S\arabic{table}}
\renewcommand{\theHfigure}{S\arabic{figure}}

\begin{center}
{\Large\bfseries Supplemental Material for\\[0.35em]
``Local phase-space Berry curvature and Hall transport in textured twisted bilayer graphene''}\par
\vspace{0.9em}
{\large T. Farajollahpour}\par
\vspace{0.35em}
{\small Department of Physics, Norwegian University of Science and Technology (NTNU), NO-7491 Trondheim, Norway\\
Department of Physics, Brock University, St. Catharines, Ontario L2S 3A1, Canada\\
\texttt{tohid.farajollahpour@ntnu.no}}
\end{center}
\vspace{1.2em}

\section{Covariant $4\times4$ Dirac formulation and the pure-geometric cancellation}

The starting point is a static covariant Dirac Hamiltonian written in terms of a spatial shift field $N^i(\bmr)$ and a tetrad $e_{\hat A}{}^{\mu}(\bmr)$,
\begin{equation}
\hat H_{\mathrm{min}}=\frac12\{N^i(\bmr),\Pi_i\}\one_4+\frac12\{\alpha^{\hat a}e_{\hat a}{}^{\,i}(\bmr),\Pi_i\}+m(\bmr)\beta,
\label{eq:sm_minimal_4x4_ham}
\end{equation}
with Hermitian symmetrization made explicit. Here $i,j\in\{x,y\}$ are spatial coordinate indices, $\mu,\nu\in\{0,x,y\}$ are spacetime coordinate indices, $\hat a,\hat b\in\{\hat x,\hat y\}$ are spatial local-frame indices, and $\hat A,\hat B\in\{\hat 0,\hat x,\hat y\}$ are full local Lorentz-frame indices. The matrices $\alpha^{\hat a}$ and $\beta$ are the usual Dirac matrices, satisfying $\{\alpha^{\hat a},\alpha^{\hat b}\}=2\delta^{\hat a\hat b}$, $\{\alpha^{\hat a},\beta\}=0$, and $\beta^2=\one_4$. Because the coefficients $N^i(\bmr)$ and $e_{\hat a}{}^{i}(\bmr)$ depend on position and therefore do not commute with $\Pi_i$, the anticommutators in Eq.~\eqref{eq:sm_minimal_4x4_ham} are required to make the Hamiltonian manifestly Hermitian.

The field $N^i(\bmr)$ is the spatial shift field: in Hamiltonian language it acts as a local frame velocity or kinematic tilt multiplying the identity sector, so it changes the local dispersion but does not by itself determine the spinor texture of the band eigenstates. The tetrad $e_{\hat A}{}^{\mu}(\bmr)$ maps coordinate indices to a local orthonormal frame and encodes the geometric deformation of the kinetic Dirac operator \cite{Birrell_1982}. The operator $\Pi_i$ is the Hermitian kinetic momentum along direction $i$; for the present geometric argument one may think of $\Pi_i=-\ii\partial_i$ in the absence of external gauge fields, or of the corresponding gauge-covariant form if such fields are present. The scalar $m(\bmr)$ is the Dirac mass. We write $m(\bmr)$ already in Eq.~\eqref{eq:sm_minimal_4x4_ham} because a real moir\'e texture may modulate the local gap; however, that modulation is a \emph{physical} texture dependence of the projected band data and is not part of the pure tetrad/shift cancellation established below.

To organize the band geometry, one separates the Hamiltonian into a scalar sector and a matrix-valued Dirac sector. Only the matrix-valued sector controls the eigenprojectors and therefore the Berry curvature. In particular, the shift term proportional to $N^i\one_4$ belongs to the scalar sector and does not enter the spinor $d$ vector directly. For the pure tetrad contribution, the relevant covariant $d$ vector is
\begin{equation}
d_{\hat A}=e_{\hat A}{}^{\mu}p_{\mu}.
\end{equation}
Here $p_{\mu}$ is the covector conjugate to $x^{\mu}=(t,x,y)$. In the static Hamiltonian problem the transport discussion later reduces to the spatial momenta $(p_x,p_y)$, but the covariant notation keeps the geometric structure transparent. The phrase ``pure tetrad sector'' means that all position dependence is carried only by the tetrad $e_{\hat A}{}^{\mu}(\bmr)$ (together with the scalar shift field $N^i$), while quantities such as $m$, the projected velocities, and the local mini-Dirac-point shift are held fixed.

The corresponding real-space covariant derivative is
\begin{equation}
\mathcal D_i d_{\hat A}=\partial_i d_{\hat A}+\Gamma^{\mu}{}_{\nu i}p_{\mu}\partial_{p_{\nu}}d_{\hat A}-\omega_{i\hat A}{}^{\hat B}d_{\hat B},
\label{eq:sm_cov_d_dA}
\end{equation}
where $\Gamma^{\mu}{}_{\nu i}$ is the Christoffel connection and $\omega_{i\hat A}{}^{\hat B}$ is the spin connection. The three terms in Eq.~\eqref{eq:sm_cov_d_dA} have distinct geometric roles: $\partial_i d_{\hat A}$ differentiates the explicit spatial dependence of the coefficients, the Christoffel term parallel-transports the coordinate index $\mu$, and the spin-connection term parallel-transports the local Lorentz index $\hat A$. Because $d_{\hat A}$ is linear in momentum, one has $p_{\mu}\partial_{p_{\nu}}d_{\hat A}=p_{\mu}e_{\hat A}{}^{\nu}$, so Eq.~\eqref{eq:sm_cov_d_dA} reduces to
\begin{equation}
\mathcal D_i d_{\hat A}=\Bigl(\partial_i e_{\hat A}{}^{\mu}+\Gamma^{\mu}{}_{\nu i}e_{\hat A}{}^{\nu}-\omega_{i\hat A}{}^{\hat B}e_{\hat B}{}^{\mu}\Bigr)p_{\mu}.
\end{equation}
The bracket is exactly the tetrad postulate, \mbox{$\nabla_i e_{\hat A}{}^{\mu}=0$}, namely the statement that the tetrad is covariantly constant once both the coordinate and Lorentz connections are included. Hence
\begin{equation}
\mathcal D_i d_{\hat A}=0.
\label{eq:sm_tetrad_postulate_zero}
\end{equation}

This is the central cancellation. The mixed phase-space Berry curvature of an isolated Dirac band is built from a momentum derivative and a real-space derivative of the same matrix-valued $d$ vector and can be written schematically as
\begin{equation}
\Om_{r_i p_j}\propto \bmd\cdot\bigl(\mathcal D_i\bmd\times \partial_{p_j}\bmd\bigr).
\end{equation}
Equation~\eqref{eq:sm_tetrad_postulate_zero} therefore implies that the pure tetrad/shift sector alone produces no mixed phase-space curvature. In other words, once the spin connection is kept consistently, a spatially varying local frame is not by itself enough to generate $\Om_{r_i p_j}$. Any nonzero mixed curvature must instead arise from additional \emph{physical} position dependence in the projected Dirac data: a spatially varying mass $m(\bmr)$, anisotropic velocities $v_a(\bmr)$, a valley-dependent mini-Dirac-point shift $A_a^{(\nu)}(\bmr)$, or related projected terms such as a tilt $w_a(\bmr)$. This is why the moir\'e analysis below is formulated in terms of texture-dependent projected miniband parameters rather than being attributed directly to the tetrad itself \cite{SundaramNiu1999,XiaoNiu2010,GaoYangNiu2014}.

\section{Local TBG cone from twist-angle and heterostrain textures}

\begin{figure}[t]
 \centering
    \includegraphics[width=0.8\linewidth]{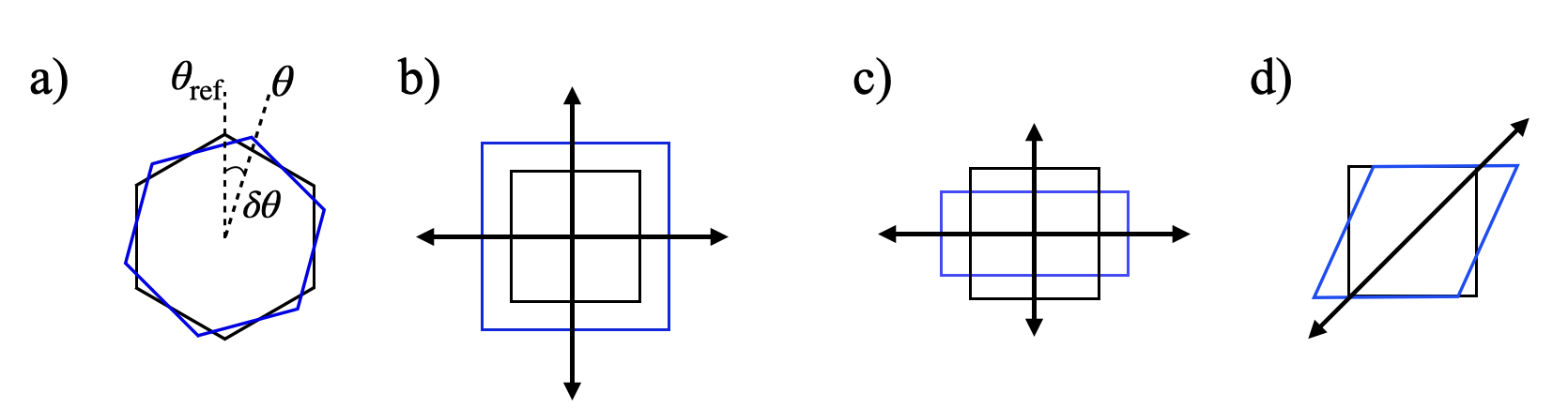}
    \caption{
Schematic illustration of the slow texture fields
$\Phi_\alpha=(\delta\theta,u_0,u_1,u_2)$ used to parameterize local moir\'e inhomogeneity.
(a) Local twist-angle deviation: the blue layer is rotated by an angle $\theta$ relative to the reference orientation $\theta_{\rm ref}$, so that $\delta\theta=\theta-\theta_{\rm ref}$.
(b) Isotropic dilation $u_0=\tfrac12(u^-_{xx}+u^-_{yy})$, corresponding to the trace part of the layer-relative strain tensor.
(c) Traceless anisotropic shear
$u_1=\tfrac12(u^-_{xx}-u^-_{yy})$, which stretches one principal axis while compressing the orthogonal one.
(d) Off-diagonal shear $u_2=u^-_{xy}$, shown as a simple shear deformation.
Black outlines denote the reference local configuration, while blue outlines denote the locally deformed one.
}
       \label{fig1-sup}
\end{figure}

In twisted bilayer graphene (TBG) the slow spatial fields are the local twist angle $\theta(\bmr)$ and the layer-relative symmetric heterostrain tensor $u_{ij}^-(\bmr)$. The superscript ``$-$'' means relative between the two layers: if $u_{ij}^{(1)}$ and $u_{ij}^{(2)}$ denote the symmetric strain tensors of the individual graphene sheets, then one may define
\[
u_{ij}^-(\bmr)=\tfrac12\bigl(u_{ij}^{(1)}(\bmr)-u_{ij}^{(2)}(\bmr)\bigr).
\]
A common strain shared by both layers mainly deforms the sample globally, whereas the \emph{relative} deformation changes the local moir\'e mismatch and therefore changes the local miniband structure. The antisymmetric part of the relative deformation is absorbed into the local twist angle, so the independent slow fields are the scalar angle $\theta(\bmr)$ and the symmetric tensor $u_{ij}^-(\bmr)$. We also define the deviation from a chosen reference twist by $\delta\theta(\bmr)=\theta(\bmr)-\theta_{\rm ref}$ (See Fig.~\ref{fig1-sup} for more details) \cite{Balents2019,Huder2018,Mesple2021,Kazmierczak2021}.

The local-moir\'e approximation assumes that these fields vary on a length scale $L_{\mathrm{tex}}$ much larger than the local moir\'e period $L_{\mathrm M}$,
\begin{equation}
L_{\mathrm{tex}}\gg L_{\mathrm M}.
\end{equation}
This separation of scales is what makes a local reduction possible. On distances large compared with $L_{\mathrm M}$ but still small compared with $L_{\mathrm{tex}}$, an electronic wave packet sees a moir\'e pattern that is nearly uniform. One may therefore freeze the texture at a given point $\bmr$, solve a local continuum-model problem there, and then allow the resulting band parameters to vary slowly from point to point. The entire supplement is written in this adiabatic local-moir\'e sense: the texture is slow enough that one first constructs a \emph{local} mini-Dirac cone and then studies how its coefficients depend on $\bmr$ \cite{BistritzerMacDonald2011,NamKoshino2017,CarrExact2019,Balents2019,Yoo2019}.

The relative strain is conveniently decomposed into one isotropic component and two traceless shear components,
\begin{equation}
 u_0=\frac12(u^-_{xx}+u^-_{yy}),\qquad
 u_1=\frac12(u^-_{xx}-u^-_{yy}),\qquad
 u_2=u^-_{xy},
\label{eq:sm_strain_invariants}
\end{equation}
so that the full symmetric relative-strain tensor can be reconstructed as
\[
u^-(\bmr)=
\begin{pmatrix}
 u_0+u_1 & u_2\\
 u_2 & u_0-u_1
\end{pmatrix}.
\]
Here $u_0$ is the layer-relative dilation, while $(u_1,u_2)$ are the two independent shear or nematic components. These are the natural symmetry-adapted variables because they separate an isotropic change of the local moir\'e scale from anisotropic distortions of the moir\'e pattern.

The local moir\'e reciprocal vectors are controlled by the relative deformation matrix
\begin{equation}
\mathcal M(\bmr)=\theta(\bmr)J-u^-(\bmr),
\qquad
J=\begin{pmatrix}0&-1\\1&0\end{pmatrix}.
\end{equation}
The matrix $J$ is the generator of an infinitesimal in-plane rotation, so the term $\theta J$ represents the local rotational mismatch of the two layers. The subtraction of $u^-$ accounts for the local strain mismatch. In the continuum description the matrix $\mathcal M(\bmr)$ controls the local moir\'e wave vectors, and therefore the local moir\'e wavelength and anisotropy. Stated differently, $\theta(\bmr)$ and $u^-_{ij}(\bmr)$ are the truly fundamental slow fields; all of the effective local-cone quantities introduced below are projected outputs of a local continuum model built from $\mathcal M(\bmr)$ \cite{Balents2019}.

Near an isolated valley-resolved mini-Dirac point of that local miniband structure, the most general linear two-band Hamiltonian can be written schematically as
\[
H^{\rm lin}_{\nu}=\varepsilon_0\one_2+w_iq_i\one_2+\sum_{a=x,y}\mathcal V^{(\nu)}_{ai}q_i\sigma_a+m\sigma_z,
\]
where $\sigma_x,\sigma_y,\sigma_z$ are Pauli matrices acting in the \emph{projected local two-state mini-Dirac subspace} and $\one_2$ is the identity in that subspace. The real $2\times 2$ matrix $\mathcal V^{(\nu)}_{ai}$ contains the linear velocity couplings between momentum and pseudospin. By a local rotation of the momentum axes, together with a corresponding rotation in the projected pseudospin basis, one may diagonalize this velocity matrix. After that local linear reduction the Hamiltonian takes the form quoted in the main text,
\begin{equation}
H_{\nu}(\bmr,\bmp)=\varepsilon_0\one_2+\bm w\!\cdot\!\bmq\,\one_2+\nu v_x q_x\sigma_x+v_y q_y\sigma_y+m\sigma_z,
\qquad q_i=p_i-A_i^{(\nu)}(\bmr).
\label{eq:sm_local_cone}
\end{equation}
The valley index $\nu=\pm$ labels the two mini-Dirac cones related by time reversal; the factor of $\nu$ in front of $v_xq_x\sigma_x$ is one convenient convention for encoding the opposite chirality of the two valleys.
The scalar $\varepsilon_0(\bmr)$ is the local energy offset of the projected cone; it shifts the local band center and therefore the relation between the global chemical potential and the local Fermi energy, but it does not affect the Berry curvature directly. The vector $\bm A^{(\nu)}(\bmr)$ specifies the location of the local mini-Dirac point in momentum space. Accordingly, $\bm q=\bmp-\bm A^{(\nu)}$ is the crystal momentum measured \emph{relative to the local cone center} rather than relative to a fixed global origin. The coefficients $v_x(\bmr)$ and $v_y(\bmr)$ are the principal-axis Dirac velocities of the local cone. The scalar $m(\bmr)$ is the local mini-Dirac mass: if $m\neq0$, the direct gap at the cone center $\bm q=0$ is $2|m|$. At the projected level this mass summarizes whatever local symmetry breaking or local hybridization opens the gap of the mini-Dirac point. Finally, $\bm w(\bmr)=(w_x,w_y)$ is an effective tilt of the cone. Because $\varepsilon_0$ and $\bm w\cdot\bm q$ multiply the identity, they do not enter the local eigenvectors and hence do not appear directly in the Berry-curvature formulas. The tilt is nevertheless essential in transport because it distorts the Fermi contour and can therefore generate a nonzero Berry-curvature dipole.

Equation~\eqref{eq:sm_local_cone} should also be read carefully as a statement about \emph{projection}. None of the coefficients $\varepsilon_0$, $\bm w$, $v_x$, $v_y$, $m$, or $\bm A^{(\nu)}$ is introduced by hand as an independent microscopic field. Rather, these are the local parameters extracted from the underlying TBG continuum problem after one projects to an isolated local mini-Dirac doublet and expands to first order about the corresponding local band crossing \cite{BistritzerMacDonald2011,CarrExact2019,Balents2019}. 

Every coefficient is a smooth function of the slow fields,
\begin{equation}
X(\bmr)=X_{\mathrm{ref}}+X_{\theta}\,\delta\theta(\bmr)+X_{u_0}u_0(\bmr)+X_{u_1}u_1(\bmr)+X_{u_2}u_2(\bmr)+\cdots,
\label{eq:sm_texture_expansion}
\end{equation}
for $X\in\{\varepsilon_0,w_x,w_y,v_x,v_y,m,A_x^{(\nu)},A_y^{(\nu)}\}$. Here $X_{\mathrm{ref}}$ is the value of the given coefficient in a chosen reference local structure, and $X_{\theta}$, $X_{u_0}$, $X_{u_1}$, and $X_{u_2}$ are the corresponding local susceptibilities. The omitted terms may contain higher-order powers such as $(\delta\theta)^2$, $u_au_b$, or mixed couplings like $\delta\theta\,u_a$. Equation~\eqref{eq:sm_texture_expansion} is therefore not a new dynamical assumption; it is simply the Taylor expansion of a smooth local band parameter in the slowly varying texture fields \cite{Huder2018,Mesple2021,Kazmierczak2021,Yu2024TBGTextures,KangVafek2025}.

In the absence of symmetry lowering one may have $\bm w=0$ to leading order, so the local cone is untilted. A nonzero effective tilt may instead be generated by heterostrain, by substrate alignment, or by other perturbations that lower the local symmetry. Likewise, in a time-reversal-symmetric two-valley problem the local cone shift is often valley odd, so that $A_i^{(-)}=-A_i^{(+)}$ in the minimal setting used later. The Hall-bar specialization in the main text makes these transformation properties even more concrete by keeping only the mirror-compatible one-dimensional texture and the valley-odd components relevant for transport.

A final technical point concerns the scalar term generated when the tilt and the cone shift are both present. Since $\bm w\cdot\bm q=\bm w\cdot\bmp-\bm w\cdot\bm A^{(\nu)}$, part of the nominal tilt actually behaves as a local scalar energy shift. In the Hall-bar reduction used later, where only $A_x$ and $w_x$ survive, the valley-even combination $-w_xA_x$ is absorbed into
\[
\varepsilon_{\rm sc}(\bmr)=\varepsilon_0(\bmr)-w_x(\bmr)A_x(\bmr),
\]
which is the local scalar contribution entering the electrochemical-equilibrium discussion. The remaining momentum-odd term $w_xp_x$ is the genuine tilt that reshapes the local Fermi surface.

The texture fields $(\theta,u^-_{ij})$ first generate a \emph{local moir\'e environment}; that local environment defines a projected mini-Dirac cone with parameters $(\varepsilon_0,\bm w,v_x,v_y,m,\bm A^{(\nu)})$; and it is the spatial dependence of those projected parameters that enters the Berry geometry. In particular, mixed phase-space curvature appears only if at least one of the genuinely vectorial Dirac data $m(\bmr)$, $v_a(\bmr)$, or $A_a^{(\nu)}(\bmr)$ varies across the sample. If all of those quantities were spatially constant, the cone would still have an ordinary momentum-space Berry curvature, but no mixed curvature.

\section{Projector reduction and TBG Berry curvatures}

For the local valley-resolved cone in Eq.~\eqref{eq:sm_local_cone}, it is useful to separate the part proportional to the identity from the part proportional to the Pauli matrices,
\begin{equation}
H_{\nu}(\bmr,\bmp)=\bigl[\varepsilon_0(\bmr)+\bm w(\bmr)\!\cdot\!\bmq\bigr]\one_2+\bmd(\bmr,\bmp)\cdot\bms,
\qquad \bms=(\sigma_x,\sigma_y,\sigma_z).
\end{equation}
The first bracket shifts the local dispersion but does not affect the band eigenvectors, whereas the second term fixes the local pseudospin texture and therefore completely determines the Berry geometry. In the present anisotropic mini-Dirac cone we write
\begin{equation}
\bmd=(d_1,d_2,d_3)=(\nu v_xq_x,\,v_yq_y,\,m),
\qquad
\varepsilon=\sqrt{d_1^2+d_2^2+d_3^2}.
\end{equation}
Here $\nu=\pm$ labels the two graphene valleys, $q_i=p_i-A_i^{(\nu)}(\bmr)$ is the momentum measured from the local valley-dependent mini-Dirac point, $v_x$ and $v_y$ are the local anisotropic Dirac velocities, and $m$ is the local Dirac mass (half the direct gap at the local cone center). The positive quantity $\varepsilon$ is the magnitude of the $d$ vector, so the two local band energies are
\begin{equation}
E_{\lambda,\nu}(\bmr,\bmp)=\varepsilon_0(\bmr)+\bm w(\bmr)\!\cdot\!\bmq+\lambda\,\varepsilon(\bmr,\bmp),
\qquad \lambda=\pm,
\end{equation}
with $\lambda=+$ for the conduction band and $\lambda=-$ for the valence band. Because the tilt term $\bm w\cdot\bmq$ multiplies the identity, it changes the local Fermi contour but does not enter the eigenprojectors or the Berry curvatures directly. All Berry-curvature formulas in this section therefore depend only on $\bmd$.

It is also useful to state the derivative convention explicitly. In the phase-space curvature tensor, indices $M$ and $N$ may denote either position coordinates $(r_x,r_y)$ or momentum coordinates $(p_x,p_y)$. A derivative $\partial_{r_i}$ is taken at fixed canonical momentum $\bmp$, whereas $\partial_{p_i}$ is taken at fixed position $\bmr$. Since $q_i=p_i-A_i^{(\nu)}(\bmr)$, a real-space derivative acts not only on the explicit spatial dependence of $m$ and $v_a$, but also on the local mini-Dirac-point shift:
\begin{equation}
\partial_i q_x=-\partial_i A_x^{(\nu)},
\qquad
\partial_i q_y=-\partial_i A_y^{(\nu)}.
\label{eq:sm_q_derivatives}
\end{equation}
This is the origin of the pseudogauge terms in the mixed-curvature formulas below.

\subsection{Projector formula for an isolated band}
For a generic two-level Hamiltonian $H=\varepsilon_{\rm id}\one_2+\bmd\cdot\bms$, the normalized pseudospin vector is $\hat{\bmd}=\bmd/\varepsilon$. The projector onto the isolated band $\lambda=\pm$ is
\begin{equation}
P_{\lambda,\nu}=\frac12\bigl(\one_2+\lambda\,\hat{\bmd}\cdot\bms\bigr),
\qquad
\hat{\bmd}=\frac{\bmd}{\varepsilon}.
\end{equation}
This projector is the operator that selects the local Bloch spinor of the chosen band at fixed $(\bmr,\bmp)$. The Berry curvature of that isolated band can be written in a manifestly gauge-invariant projector form \cite{SundaramNiu1999,XiaoNiu2010},
\begin{equation}
\mathcal F_{MN}^{(\lambda,\nu)}=\ii P_{\lambda,\nu}[\partial_M P_{\lambda,\nu},\partial_N P_{\lambda,\nu}]P_{\lambda,\nu},
\end{equation}
where $M,N\in\{r_x,r_y,p_x,p_y\}$ label phase-space coordinates. 

Because the band is one-dimensional after projection, the operator-valued curvature is proportional to the projector itself,
\begin{equation}
\mathcal F_{MN}^{(\lambda,\nu)}
=\Om_{MN}^{(\lambda,\nu)}P_{\lambda,\nu},
\end{equation}
where $\Om_{MN}^{(\lambda,\nu)}$ is the scalar (Abelian) Berry curvature associated with that band. Using $(\hat{\bmd}\cdot\bms)^2=\one_2$ and the Pauli-matrix commutator $[\sigma_a,\sigma_b]=2\ii\epsilon_{abc}\sigma_c$, one finds after standard algebra
\begin{equation}
\Om_{MN}^{(\lambda,\nu)}=-\frac{\lambda}{2\varepsilon^3}\,\bmd\cdot\bigl(\partial_M\bmd\times\partial_N\bmd\bigr).
\label{eq:sm_projector_formula}
\end{equation}
Equation~\eqref{eq:sm_projector_formula} is the master formula used throughout the paper. It shows that the Berry geometry is the geometry of the mapping from phase space $(\bmr,\bmp)$ into the pseudospin vector $\bmd$. Whenever the pseudospin texture twists as a function of position or momentum, a Berry curvature is generated.

\subsection{Momentum-space curvature}
The ordinary momentum-space curvature follows from differentiating $\bmd$ with respect to $p_x$ and $p_y$ at fixed position. Since only $q_x$ depends on $p_x$, only $q_y$ depends on $p_y$, and only the mass component $d_3=m$ contributes to the scalar triple product in Eq.~\eqref{eq:sm_projector_formula}, the result is
\begin{equation}
\Om_{p_xp_y}^{(\lambda,\nu)}=-\frac{\lambda\nu v_xv_y m}{2\varepsilon^3}.
\label{eq:sm_pp_curvature}
\end{equation}
This is the familiar Berry curvature of a massive anisotropic Dirac cone \cite{XiaoNiu2010}. Its sign is controlled by the band index $\lambda$, the valley index $\nu$, and the local mass $m$. The curvature is largest near the local band edge, where $\varepsilon$ is smallest, and it decays as $\varepsilon^{-3}$ away from the mini-Dirac point.

\subsection{Mixed curvature}
The mixed phase-space curvature is obtained by taking one derivative with respect to position and one with respect to momentum. Because $q_i$ depends on position through $A_i^{(\nu)}(\bmr)$, the real-space derivative of the $d$ vector contains both genuine parameter gradients and pseudogauge gradients. Equation~\eqref{eq:sm_projector_formula} then yields
\begin{equation}
\Om_{r_i p_x}^{(\lambda,\nu)}
=-\frac{\lambda\nu v_xv_y}{2\varepsilon^3}
\Bigl[q_y\bigl(\partial_i m-m\partial_i\ln v_y\bigr)+m\partial_iA_y^{(\nu)}\Bigr].
\label{eq:sm_mixed_px}
\end{equation}
and 
\begin{equation}
\Om_{r_i p_y}^{(\lambda,\nu)}
=\phantom{-}\frac{\lambda\nu v_xv_y}{2\varepsilon^3}
\Bigl[q_x\bigl(\partial_i m-m\partial_i\ln v_x\bigr)+m\partial_iA_x^{(\nu)}\Bigr].
\label{eq:sm_mixed_py}
\end{equation}
These two equations make the structure of the mixed curvature transparent. There are two independent sources. The first is the gradient combination $\partial_i m-m\partial_i\ln v_a$,  which measures how the local mass changes relative to the local velocity scale. The second is the direct pseudogauge contribution $m\partial_iA_a^{(\nu)}$, which survives even in a local patch where $m$ and the velocities are approximately constant. This latter limit is the one used later in the Hall-bar calculation.

The texture origin of these derivatives becomes explicit by applying the chain rule to any local cone parameter $X\in\{m,v_x,v_y,A_x^{(\nu)},A_y^{(\nu)}\}$,
\begin{equation}
\partial_iX=\frac{\partial X}{\partial\theta}\,\partial_i\theta+
\sum_{\chi\in\{u_0,u_1,u_2\}}\frac{\partial X}{\partial\chi}\,\partial_i\chi+\cdots,
\label{eq:sm_chain_rule}
\end{equation}
where the derivatives of $X$ with respect to the slow fields are the local susceptibilities introduced in Eq.~\eqref{eq:sm_texture_expansion}. Equation~\eqref{eq:sm_chain_rule} shows explicitly how gradients of the twist angle and heterostrain feed into the mixed Berry curvature: a nonzero $\partial_i\theta$ or $\partial_i u^-_{jk}$ produces $\Om_{r_ip_j}$ whenever the local projected cone parameters respond to those fields.

The effective tilt $\bm w$ does not appear in Eqs.~\eqref{eq:sm_mixed_px} and \eqref{eq:sm_mixed_py} because it multiplies the identity operator and therefore does not rotate the local pseudospin vector. In this sense mixed curvature and tilt play complementary roles in the later transport theory: the former controls the linear phase-space correction, whereas the latter is needed to generate a Berry-curvature dipole and hence a dc nonlinear Hall response.

\subsection{Real-space curvature}
For completeness, we also comment on the purely real-space Berry curvature $\Om_{r_xr_y}^{(\lambda,\nu)}$. In the full problem it contains products of gradients of $m$, $v_x$, $v_y$, and $A_i^{(\nu)}$, so the general expression is algebraically longer than the mixed-curvature formulas and is not needed for the transport calculation in the main text. The limit relevant for the minimal pseudogauge discussion is particularly simple: treat $m$, $v_x$, and $v_y$ as locally constant and allow only the mini-Dirac-point shift $A_i^{(\nu)}(\bmr)$ to vary.  Equation~\eqref{eq:sm_projector_formula} gives
\begin{equation}
\Om_{r_xr_y}^{(\lambda,\nu)}
=-\frac{\lambda\nu v_xv_y m}{2\varepsilon^3}
\Bigl[(\partial_xA_x^{(\nu)})(\partial_yA_y^{(\nu)})-(\partial_xA_y^{(\nu)})(\partial_yA_x^{(\nu)})\Bigr].
\label{eq:sm_rr_curvature}
\end{equation}
Two features are worth emphasizing. First, the pure-$A$ contribution is quadratic in spatial gradients. It is therefore not proportional to the curl-like pseudomagnetic combination $\partial_xA_y^{(\nu)}-\partial_yA_x^{(\nu)}$ that appears in the mixed-curvature Hall-bar limit. Second, in the one-dimensional Hall-bar specialization used later,
\begin{equation}
A_x=A_x(y),\qquad A_y=0,
\end{equation}
Eq.~\eqref{eq:sm_rr_curvature} vanishes identically because either an $x$ derivative or an $A_y$ factor is always missing. Thus the relevant geometric correction in the transport calculation is not a real-space Berry curvature but the mixed phase-space curvature derived above. This is why the pseudomagnetic field enters the longitudinal transport through $\Om_{r_ip_j}$ rather than through $\Om_{r_xr_y}$.

\section{Mirror analysis for the Hall-bar geometry}
The physical sample is  a two-dimensional TBG device, but the slow moir\'e texture is assumed to vary only across the Hall bar, namely along the coordinate $y$. The orthogonal direction $x$ is the direction along the bar, so it is the natural axis for a longitudinal drive. In the minimal mirror-preserving reduction we retain only the valley-odd displacement of the local mini-Dirac point along $x$ and set $A_y=0$. The local valley-resolved Hamiltonian then takes the form
\begin{equation}
H_{\nu}=\varepsilon_0\one_2+\nu w_x(y)q_x\one_2+\nu v_x q_x\sigma_x+v_y q_y\sigma_y+m\sigma_z,
\qquad q_x=p_x-\nu A_x(y),\quad q_y=p_y.
\label{eq:sm_transport_setup}
\end{equation}
Here $\nu=\pm$ labels the two graphene valleys, $\varepsilon_0(y)$ is a local scalar shift of the cone, $w_x(y)$ is an effective tilt along the bar, $v_x$ and $v_y$ are the local Dirac velocities, and $m$ is the local Dirac mass, so the direct band gap at the cone is $2|m|$. The momenta $q_i$ are measured relative to the local valley-dependent mini-Dirac point. In particular, the quantity $A_x(y)$ acts as a valley-odd pseudogauge potential: when it varies with $y$, its derivative $\partial_yA_x$ will later play the role of the local pseudomagnetic field that enters the mixed-curvature correction \cite{AlEzzi2025}.

At this stage we are not yet computing a transport coefficient; we are only asking which tensor components are allowed by symmetry before any dynamics is inserted. The relevant symmetry is the physical mirror of the full two-valley Hall bar,
\begin{equation}
M_x:(x,y)\mapsto(-x,y),
\end{equation}
which flips the coordinate along the bar and leaves the textured direction unchanged. The symmetry statement applies to the valley-summed physical current. A single valley need not be separately invariant under the mirror, but once both valleys are included the observable response of the device must respect this spatial symmetry. Under $M_x$ the $x$ component of any in-plane polar vector is odd under the mirror, whereas the $y$ component is even.

This immediately determines how the measured current density $\bm j=(j_x,j_y)$ and the applied electric field $\bm E=(E_x,E_y)$ transform:
\begin{equation}
j_x\to-j_x,
\qquad
j_y\to j_y,
\qquad
E_x\to-E_x,
\qquad
E_y\to E_y.
\end{equation}
It is convenient to encode these parities as
\begin{equation}
s_x=-1,\qquad s_y=+1,
\end{equation}
so that any in-plane polar vector $V_i$ transforms as $V_i\to s_iV_i$. We can now apply this bookkeeping directly to the linear and nonlinear transport tensors. For linear response one writes
\begin{equation}
j_i=\sigma_{ij}E_j.
\end{equation}
Using $j_i\to s_ij_i$ and $E_j\to s_jE_j$, one finds
\begin{equation}
\sigma_{ij}=s_is_j\sigma_{ij}.
\label{eq:sm_linear_mirror_constraint}
\end{equation}
For $s_is_j=+1$ the coefficient is mirror even and may be nonzero. For $s_is_j=-1$ the coefficient is mirror odd and must vanish in a mirror-symmetric device. Therefore the longitudinal conductivities $\sigma_{xx}$ and $\sigma_{yy}$ are allowed, but the off-diagonal coefficients are forbidden
\begin{equation}
\sigma_{xy}=\sigma_{yx}=0.
\end{equation}
This is the symmetry reason that the Hall-bar geometry used here cannot exhibit a linear Hall response as long as the residual mirror and the two-valley structure are preserved. Any nonzero linear Hall signal in such a device would therefore indicate additional symmetry breaking beyond the minimal setup.

For the quadratic response one writes
\begin{equation}
j_i=\chi_{ilm}E_lE_m.
\end{equation}
Because the electric-field factors commute, the response tensor may be taken symmetric in its last two indices,
\begin{equation}
\chi_{ilm}=\chi_{iml}.
\end{equation}
Applying the mirror to the quadratic constitutive relation gives the constraint
\begin{equation}
\chi_{ilm}=s_is_ls_m\chi_{ilm}.
\label{eq:sm_quadratic_mirror_constraint}
\end{equation}
A quadratic coefficient is allowed only if the product of the parities of its three indices is $+1$. In two dimensions this removes all components with an odd number of $x$ indices. Explicitly, $\chi_{xxx}$, $\chi_{xyy}$, $\chi_{yxy}$, and $\chi_{yyx}$ are mirror odd and therefore vanish, while the mirror-even components are
\begin{equation}
\chi_{yxx},\qquad \chi_{xxy}=\chi_{xyx},\qquad \chi_{yyy}.
\label{eq:sm_mirror_allowed_chi}
\end{equation}
The coefficient $\chi_{yxx}$ describes the transverse second-order current generated by a longitudinal drive $E_x$; this is the nonlinear Hall channel emphasized in the main text. The pair $\chi_{xxy}=\chi_{xyx}$ describes a mixed response that requires both field components to be present simultaneously. The coefficient $\chi_{yyy}$ would represent a quadratic correction to the longitudinal current along the textured direction.

Mirror symmetry alone does \emph{not} force $\chi_{yyy}$ to vanish. All three indices are $y$, so the coefficient is even under $M_x$. The stronger result $\chi_{yyy}=0$ found later in the explicit Boltzmann calculation is therefore dynamical rather than kinematical. In the minimal Hall-bar cone the Berry-curvature dipole has only an $x$ component, while $D_y=0$ by parity of the underlying momentum integral. Consequently the only nonlinear Hall coefficient that survives for a pure longitudinal drive $E=(E_x,0)$ is $\chi_{yxx}$.

\section{Semiclassical equations and Boltzmann solution}
We now spell out in more detail how the local Hall-bar cone is turned into transport formulas. Throughout this section we work with the conduction band $\lambda=+$ in a single local patch of the textured sample. The valley index is $\nu=\pm$, the momenta measured relative to the local valley-dependent mini-Dirac point are
\begin{equation}
q_x=p_x-\nu A_x(y),
\qquad
q_y=p_y,
\end{equation}
and the positive pseudospin-sector energy is
\begin{equation}
\varepsilon(\bmp;y)=\sqrt{v_x^2q_x^2+v_y^2q_y^2+m^2}.
\end{equation}
In the present one-dimensional Hall-bar geometry, $\calB_s(y)\equiv \partial_yA_x(y)$, so that $\calB_s$ is the pseudomagnetic field generated by the texture gradient.

Substituting this specialization into the general projector formulas derived above gives, for the conduction band,
\begin{equation}
\Om_{p_xp_y}^{(+,\nu)}=-\frac{\nu v_xv_y m}{2\varepsilon^3},
\qquad
\Om_{yp_y}^{(+,\nu)}=\frac{v_xv_y m}{2\varepsilon^3}\,\calB_s,
\label{eq:sm_selected_curvatures}
\end{equation}
with $\Om_{yp_x}^{(+,\nu)}=0$ in the local patch where $\partial_y m=\partial_y v_x=\partial_y v_y=0$. The momentum-space curvature $\Om_{p_xp_y}^{(+,\nu)}$ is odd in valley, as expected for a time-reversal-symmetric two-valley Dirac problem. By contrast, the mixed curvature $\Om_{yp_y}^{(+,\nu)}$ is valley even in the Hall-bar geometry, because the explicit factor of $\nu$ in Eq.~\eqref{eq:sm_mixed_py} is canceled by the derivative of the valley-odd cone shift, $\partial_yA_x^{(\nu)}=\nu\calB_s$. This valley-even character is what allows the mixed-curvature correction to survive after summing over valleys.

The local conduction-band energy can be rewritten as
\begin{equation}
E_{+,\nu}(y,\bmp)=\varepsilon_{\rm sc}(y)+\nu w_x(y)p_x+\varepsilon[q_x(y),q_y;y],
\qquad
\varepsilon_{\rm sc}(y)\equiv \varepsilon_0(y)-w_x(y)A_x(y).
\label{eq:sm_local_energy}
\end{equation}
Here $\varepsilon_0(y)$ is the local scalar shift of the cone and $\varepsilon_{\rm sc}(y)$ is the valley-even scalar combination that should be absorbed into the local electrochemical equilibrium. In the main-text notation, this same scalar sector shifts the global electrochemical potential according to $\mu_{\rm loc}(y)=\mu_{\rm glob}-\varepsilon_{\rm sc}(y)$. The remaining term $\nu w_xp_x$ is valley odd and does not modify the Berry curvature directly; instead it distorts the local Fermi contour and later generates the Berry-curvature dipole.

The distribution function in valley $\nu$ is denoted $f_\nu(\bmr,\bmp)$. In the steady relaxation-time approximation the natural local reference state is
\begin{equation}
f_\nu^{\rm loc}(\bmr,\bmp)=f_0\!\bigl(E_{+,\nu}(\bmr,\bmp)-\mu_{\rm loc}(y)\bigr),
\end{equation}
where $f_0(\xi)=1/(e^{\beta\xi}+1)$ is the Fermi function, $\mu_{\rm loc}(y)$ is the local electrochemical potential, and $\tau$ is the transport relaxation time \cite{Sharma2021Transport}. 

The steady kinetic equation is
\begin{equation}
\dot r_i\partial_{r_i}f_\nu+\dot p_i\partial_{p_i}f_\nu=-\frac{f_\nu-f_\nu^{\rm loc}}{\tau}.
\label{eq:sm_boltzmann_general}
\end{equation}
Equation~\eqref{eq:sm_boltzmann_general} makes explicit which terms are present in the full spatially varying problem. First, because the local band energy depends on $y$, the force $\dot p_y$ contains not only the external electric force but also a scalar contribution $-\partial_yE_{+,\nu}$. Second, the left-hand side contains the real-space drift term $\dot r_i\partial_{r_i}f_\nu$. These are the terms that would have to be retained in a globally controlled device-scale treatment.

The closed analytic formulas used in the main text follow from a more limited but controlled \emph{local-frame} approximation. One chooses a patch that is small on the texture scale $L_{\rm tex}$ but large on microscopic scales, treats $\mu_{\rm loc}$, $w_x$, $m$, $v_x$, and $v_y$ as constant within that patch, and retains only the pseudogauge gradient $\partial_yA_x$. Concretely, we impose
\begin{equation}
\partial_y\mu_{\rm loc}=0,
\qquad
\partial_y w_x=\partial_y m=\partial_y v_x=\partial_y v_y=0,
\end{equation}
while keeping $\calB_s=\partial_yA_x\neq 0$. In this reduction the scalar-force contribution is absorbed into the definition of $f_\nu^{\rm loc}$, the nonequilibrium correction is treated as locally translationally invariant, and the transport problem reduces to a momentum-space calculation controlled by the mixed curvature.

It is crucial to use the same phase-space symplectic structure for both the equations of motion and the density of states. We order phase-space coordinates as
\begin{equation}
\xi_a=(x,y,p_x,p_y),
\end{equation}
and keep only the nonzero Berry-curvature components of the local Hall-bar patch, namely $\Om_{p_xp_y}$ and $\Om_{yp_y}$. To linear order in these curvatures the symplectic matrix is
\begin{equation}
\omega_{ab}=
\begin{pmatrix}
0 & 0 & -1 & 0\\
0 & 0 & 0 & -1+\Om_{yp_y}\\
1 & 0 & 0 & \Om_{p_xp_y}\\
0 & 1-\Om_{yp_y} & -\Om_{p_xp_y} & 0
\end{pmatrix}.
\label{eq:sm_symplectic_matrix}
\end{equation}
The off-diagonal $\Om_{p_xp_y}$ entries encode the ordinary momentum-space Berry curvature, while the deviation of the $(y,p_y)$ block from the canonical form is the mixed-curvature correction. In a more general local patch an additional scalar-force term $-\partial_yE_{+,\nu}$ would appear in $\dot p_y$ after inverting $\omega_{ab}$. Within the local-frame approximation described above, however, inverting Eq.~\eqref{eq:sm_symplectic_matrix} gives the reduced semiclassical equations
\begin{align}
\dot p_x&=-eE_x,\notag\\
\dot p_y&=-e(1+\Om_{yp_y})E_y,\notag\\
\dot r_x&=v_x^{\mathrm g}+e\Om_{p_xp_y}E_y,\notag\\
\dot r_y&=(1+\Om_{yp_y})v_y^{\mathrm g}-e\Om_{p_xp_y}E_x,
\label{eq:sm_eom}
\end{align}
where the local group velocities are obtained by differentiating $E_{+,\nu}$ with respect to momentum 
\begin{equation}
v_x^{\mathrm g}=\partial_{p_x}E_{+,\nu}=\nu w_x+\frac{v_x^2q_x}{\varepsilon},
\qquad
v_y^{\mathrm g}=\partial_{p_y}E_{+,\nu}=\frac{v_y^2q_y}{\varepsilon}.
\end{equation}
Thus the mixed curvature renormalizes the $y$-sector kinematics, whereas the anomalous-velocity term proportional to $\Om_{p_xp_y}$ produces the usual transverse Berry correction. The same symplectic matrix also fixes the invariant phase-space measure \cite{SundaramNiu1999,XiaoNiu2010,GaoYangNiu2014},
\begin{equation}
\mathcal D(\bmp)=\sqrt{\det\omega}=1-\Om_{yp_y}+O(\Om^2).
\label{eq:sm_phase_space_measure}
\end{equation}

With the local-frame reduction in place, the kinetic equation becomes purely momentum-space
\begin{equation}
\dot p_i\,\partial_{p_i}f_{\nu}=-\frac{f_{\nu}-f_{\nu}^{\rm loc}}{\tau}.
\label{eq:sm_boltzmann}
\end{equation}
We expand around the local equilibrium state as
\begin{equation}
f_{\nu}=f_{\nu}^{\rm loc}+\delta f_{\nu}^{(1)}+\delta f_{\nu}^{(2)}+\cdots.
\end{equation}
To first order in the electric field,
\begin{equation}
\delta f_{\nu}^{(1)}=-\tau\,\dot p_i\,\partial_{p_i}f_{\nu}^{\rm loc}
=e\tau\,\mathcal E_i\,\partial_{p_i}f_0,
\end{equation}
and iterating once more gives
\begin{equation}
\delta f_{\nu}^{(2)}=-\tau\,\dot p_i\,\partial_{p_i}\delta f_{\nu}^{(1)}
=e^2\tau^2\,\mathcal E_i\mathcal E_j\,\partial_{p_i}\partial_{p_j}f_0+\cdots,
\label{eq:sm_fexpansion}
\end{equation}
where the effective driving fields are
\begin{equation}
\mathcal E_x=E_x,
\qquad
\mathcal E_y=(1+\Om_{yp_y})E_y.
\end{equation}

Finally, the physical current density is obtained by summing the band velocity over both valleys and over spin,
\begin{equation}
j_i=-g_s e\sum_{\nu=\pm}\int\frac{\dd^2p}{(2\pi)^2}\,\mathcal D(\bmp)\,\dot r_i f_{\nu}.
\label{eq:sm_current}
\end{equation}
Here $g_s=2$ is the spin degeneracy, while the valley sum is written explicitly because different ingredients behave differently under $\nu\to-\nu$: $\Om_{p_xp_y}$ is valley odd, $\Om_{yp_y}$ is valley even in the Hall-bar geometry, and the tilt term $\nu w_xp_x$ is again valley odd. This separation is what allows the linear Hall response to vanish while the longitudinal mixed-curvature correction survives and the nonlinear Hall response requires an additional valley-odd tilt. For notational simplicity, in the remaining sections we denote the local Fermi energy of the chosen patch simply by $\mu\equiv\mu_{\rm loc}$.

\section{Linear conductivity: Drude part and mixed-curvature correction}
We evaluate the linear conductivity of the local Hall-bar cone in the conduction band. Throughout this section $\mu\equiv\mu_{\rm loc}$ denotes the local chemical potential of the chosen patch, measured relative to the locally shifted conduction-band edge. A Fermi surface exists only for $\mu>|m|$, so the local band is metallic rather than insulating. At zero temperature the derivative of the Fermi function collapses onto the Fermi contour,
\begin{equation}
-\partial_{\varepsilon}f_0=\delta(\mu-\varepsilon),
\qquad
k_F=\sqrt{\mu^2-m^2},
\label{eq:sm_zeroT_identities}
\end{equation}
where $k_F$ is the radius of the Fermi circle in the isotropized variables introduced below. We remove the anisotropy of the Dirac cone by defining
\begin{equation}
k_x=v_xq_x,
\qquad
k_y=v_yq_y,
\qquad
\varepsilon=\sqrt{k_x^2+k_y^2+m^2},
\qquad
\dd^2q=\frac{\dd^2k}{v_xv_y}.
\label{eq:sm_isotropized_variables}
\end{equation}
In the $\bm{k}=(k_x,k_y)$ variables the equal-energy contours are circles. Writing $k_x=k\cos\phi$ and $k_y=k\sin\phi$, one finds
\begin{equation}
\int\frac{\dd^2k}{(2\pi)^2}\,F(\phi)\,\delta(\mu-\varepsilon)
=\frac{\mu}{(2\pi)^2}\int_0^{2\pi}\dd\phi\,F(\phi)\Big|_{k=k_F},
\label{eq:sm_fs_identity}
\end{equation}
for any smooth angular function $F(\phi)$. 

\subsection{Drude conductivity along $x$}

We first consider a purely longitudinal drive $E=(E_x,0)$. In the minimal Hall-bar patch the mixed curvature modifies only the $y$-sector kinematics, so the analytic $x$-channel response remains of Drude form. In addition, the tilt contribution $\nu w_x$ to the group velocity is odd in the valley index and therefore cancels in the two-valley sum. 
The linear current is therefore
\begin{align}
\sigma_{xx}
=g_sg_v e^2\tau\int\frac{\dd^2q}{(2\pi)^2}
\left(\frac{v_x^2q_x}{\varepsilon}\right)^2
\delta(\mu-\varepsilon)
=\frac{g_sg_v e^2\tau}{4\pi}\,\frac{v_x}{v_y}\,\frac{\mu^2-m^2}{\mu}.
\label{eq:sm_sigma_xx}
\end{align}
This is the anisotropic Drude conductivity of the local mini-Dirac cone. The factor $v_x/v_y$ reflects the ellipticity of the original cone in $\bmq$ space, while the factor $(\mu^2-m^2)/\mu$ is the standard massive-Dirac phase-space weight.

\subsection{Conductivity along $y$}
For a drive $E=(0,E_y)$ the calculation differs in one essential respect: the mixed curvature $\Om_{yp_y}$ enters both the equation of motion and the invariant phase-space measure. Using Eqs.~\eqref{eq:sm_eom} and \eqref{eq:sm_phase_space_measure}, the linear current along $y$ becomes
\begin{align}
\sigma_{yy}
=g_sg_v e^2\tau\int\frac{\dd^2q}{(2\pi)^2}
\bigl(1+\Om_{yp_y}\bigr)
\left(\frac{v_y^2q_y}{\varepsilon}\right)^2
\delta(\mu-\varepsilon)+O(\Om^2).
\label{eq:sm_sigma_yy_start}
\end{align}

This separates the result into an ordinary Drude part and a geometric correction,
\begin{equation}
\sigma_{yy}=\sigma_{yy}^{\mathrm D}+\delta\sigma_{yy}^{\mathrm{mix}}.
\end{equation}
The Drude contribution is obtained by dropping $\Om_{yp_y}$ in Eq.~\eqref{eq:sm_sigma_yy_start}:
\begin{align}
\sigma_{yy}^{\mathrm D}
=g_sg_v e^2\tau\int\frac{\dd^2q}{(2\pi)^2}
\left(\frac{v_y^2q_y}{\varepsilon}\right)^2\delta(\mu-\varepsilon)
=\frac{g_sg_v e^2\tau}{4\pi}\,\frac{v_y}{v_x}\,\frac{\mu^2-m^2}{\mu}.
\label{eq:sm_sigma_yy_drude}
\end{align}
This is the same massive-Dirac Drude result as along $x$, with the anisotropy inverted because the current now probes the $y$ component of the group velocity.

The mixed-curvature correction is the part linear in $\Om_{yp_y}$,
\begin{equation}
\delta\sigma_{yy}^{\mathrm{mix}}
=g_sg_v e^2\tau\int\frac{\dd^2q}{(2\pi)^2}
\Om_{yp_y}
\left(\frac{v_y^2q_y}{\varepsilon}\right)^2
\delta(\mu-\varepsilon).
\label{eq:sm_delta_sigma_yy_mix_start}
\end{equation}
On the Fermi contour the mixed Berry curvature becomes the constant
\begin{equation}
\Om_F\equiv \Om_{yp_y}(\varepsilon=\mu)=\frac{v_xv_y m\,\calB_s}{2\mu^3}.
\label{eq:sm_OmegaF_def}
\end{equation}
Because $\Om_F$ is independent of the angular coordinate $\phi$, it can be pulled out of the Fermi-surface integral in Eq.~\eqref{eq:sm_delta_sigma_yy_mix_start}. The remaining integral is exactly the Drude one, so
\begin{equation}
\delta\sigma_{yy}^{\mathrm{mix}}=\Om_F\sigma_{yy}^{\mathrm D}
=\sigma_{yy}^{\mathrm D}\,\frac{v_xv_y m\,\calB_s}{2\mu^3}.
\label{eq:sm_delta_sigma_yy_mix}
\end{equation}
The correction is proportional to the ordinary conductivity of the local cone, multiplied by a dimensionless measure of the mixed phase-space curvature at the Fermi surface.


\section{Berry-curvature dipole and nonlinear Hall coefficients}
In a time-reversal-invariant two-dimensional metal this sector is naturally organized by the ordinary momentum-space Berry curvature and its dipole over the occupied states \cite{SodemannFu2015,Ma2019NLH,Kang2019NLH,Ma2021Review,Du2021Review,Zhang2022NLH,Duan2022NLH,Huang2023INH,Chen2024Crossed}. In the present local Hall-bar cone there are two conceptually distinct quadratic contributions
\begin{equation}
j_i^{(2)}=j_{i,\mathrm D}^{(2)}+j_{i,\mathrm{BCD}}^{(2)}.
\end{equation}
The first term $j_{i,\mathrm D}^{(2)}$ comes from the ordinary group velocity multiplied by the second-order nonequilibrium correction $\delta f_\nu^{(2)}$. The second term $j_{i,\mathrm{BCD}}^{(2)}$ comes from the anomalous velocity $e\epsilon_{ij}\Omega_{p_xp_y}^{(+,\nu)}E_j$ multiplied by the first-order correction $\delta f_\nu^{(1)}$. For the present mirror-preserving minimal cone, the dc Drude-type contribution does not survive after angular integration and valley summation: the relevant integrands are either parity odd on the Fermi contour or odd under $\nu\to-\nu$. The leading dc quadratic response is therefore entirely the Berry-curvature-dipole (BCD) term
\begin{equation}
j_i^{(2)}=-g_s\sum_{\nu=\pm}e^2\epsilon_{ij}E_j\int\frac{\dd^2p}{(2\pi)^2}\,\Om_{p_xp_y}^{(+,\nu)}f_{\nu}^{(1)}.
\end{equation}
This simplification is specific to the minimal Hall-bar model and should not be interpreted as a general statement about all textured TBG devices.

To extract the leading BCD contribution, 
one may rewrite the quadratic current in the standard dipole form \cite{SodemannFu2015}
\begin{equation}
j_i^{(2)}=\chi_{ilm}^{\mathrm{BCD}}E_lE_m,
\qquad
\chi_{ilm}^{\mathrm{BCD}}=\frac{e^3\tau}{2}\bigl(\epsilon_{il}D_m+\epsilon_{im}D_l\bigr),
\label{eq:sm_chi_from_D}
\end{equation}
where
\begin{equation}
D_a=g_s\sum_{\nu=\pm}\int\frac{\dd^2p}{(2\pi)^2}\,f_0(E_{+,\nu})\partial_{p_a}\Om_{p_xp_y}^{(+,\nu)}.
\label{eq:sm_Ddef}
\end{equation}
The vector $D_a$ is the Berry-curvature dipole \cite{SodemannFu2015,Ma2021Review,Du2021Review}. It measures the first momentum-space moment of the Berry curvature over the occupied states. In an inversion-symmetric or perfectly untilted Fermi contour, opposite points contribute equally and $D_a$ vanishes. A nonzero dipole therefore requires not merely Berry curvature, but an additional asymmetry of the occupied Fermi sea.

For the $x$ component one finds
\begin{align}
D_x
&=-g_sw_x\sum_{\nu}\int\frac{\dd^2p}{(2\pi)^2}\,\nu q_x\delta(\mu-\varepsilon)\,\partial_{p_x}\Om_{p_xp_y}^{(+,\nu)}\notag\\
&=-\frac{g_sg_v\,3mw_x}{2}\int\frac{\dd^2k}{(2\pi)^2}\frac{k_x^2}{\varepsilon^5}\delta(\mu-\varepsilon),
\label{eq:sm_Dx_start}
\end{align}
where in the second line we used the isotropized variables $k_x=v_xq_x$, $k_y=v_yq_y$, the Jacobian $\dd^2p=\dd^2q=\dd^2k/(v_xv_y)$, and the fact that the explicit valley sum now produces $g_v=2$. Evaluating the remaining Fermi-surface integral gives
\begin{align}
D_x
=-\frac{g_sg_v\,3mw_x}{8\pi}\,\frac{\mu^2-m^2}{\mu^4}.
\label{eq:sm_Dx}
\end{align}
By contrast,
\begin{equation}
D_y=0,
\end{equation}
because even after the tilt expansion the full integrand remains odd under $q_y\to-q_y$. 
The tilt $w_x$ distorts the Fermi contour only along $x$, so only the $x$ component of the dipole can survive. The mirror analysis allowed $\chi_{yyy}$ on symmetry grounds, but the explicit dynamics of the minimal cone force it to vanish because there is no corresponding $D_y$.

Substituting Eq.~\eqref{eq:sm_Dx} into Eq.~\eqref{eq:sm_chi_from_D} yields the surviving nonlinear coefficients,
\begin{align}
\chi_{yxx}&=-e^3\tau D_x
=\frac{g_sg_v\,3e^3\tau mw_x}{8\pi}\,\frac{\mu^2-m^2}{\mu^4},
\label{eq:sm_chi_yxx}\\
\chi_{xxy}&=\chi_{xyx}=\frac{e^3\tau}{2}D_x
=-\frac{g_sg_v\,3e^3\tau mw_x}{16\pi}\,\frac{\mu^2-m^2}{\mu^4},
\label{eq:sm_chi_xxy}
\end{align}
with all remaining quadratic coefficients vanishing in the minimal model,
\begin{equation}
\chi_{xxx}=\chi_{xyy}=\chi_{yyx}=\chi_{yxy}=\chi_{yyy}=0.
\label{eq:sm_chi_zero}
\end{equation}

\section{Full current decomposition and surviving coefficients}

We now collect the linear and quadratic results into a single local constitutive relation for the mirror-preserving Hall-bar patch. Up to second order in the electric field,
\begin{equation}
j_i=j_i^{\mathrm D}+j_i^{\mathrm{mix}}+j_i^{\mathrm{BCD}}+O(E^3),
\end{equation}
with
\begin{align}
j_i^{\mathrm D}&=\sigma_{ij}^{\mathrm D}E_j,\notag\\
j_i^{\mathrm{mix}}&=\delta\sigma_{ij}^{\mathrm{mix}}E_j,\notag\\
j_i^{\mathrm{BCD}}&=\chi_{ilm}^{\mathrm{BCD}}E_lE_m.
\end{align}
This decomposition should be read as a hierarchy of mechanisms. The Drude sector is present already in a uniform local cone. The mixed sector is a linear-in-field correction controlled by the phase-space curvature $\Om_{yp_y}$ and therefore by the pseudogauge gradient $\calB_s=\partial_yA_x$. The BCD sector is a genuinely second-order response controlled by the ordinary momentum-space Berry curvature together with the extra Fermi-surface asymmetry induced by the valley-odd tilt $w_x$.

Because the linear Hall tensor vanishes and the only nonzero BCD coefficients are $\chi_{yxx}$ and $\chi_{xxy}=\chi_{xyx}=-\chi_{yxx}/2$, the current for a general in-plane field $E=(E_x,E_y)$ reduces to
\begin{align}
j_x&=\sigma_{xx}^{\mathrm D}E_x-\chi_{yxx}^{\mathrm{BCD}}E_xE_y,
\label{eq:sm_jx_full}\\
j_y&=\bigl(\sigma_{yy}^{\mathrm D}+\delta\sigma_{yy}^{\mathrm{mix}}\bigr)E_y+\chi_{yxx}^{\mathrm{BCD}}E_x^2.
\label{eq:sm_jy_full}
\end{align}
The crossed quadratic term in $j_x$ is simply the combination
\begin{equation}
j_x^{(2)}=\chi_{xxy}E_xE_y+\chi_{xyx}E_yE_x
=2\chi_{xxy}E_xE_y
=-\chi_{yxx}E_xE_y,
\end{equation}
while no $E_y^2$ contribution survives because $\chi_{yyy}=0$ dynamically in the minimal cone. Equations~\eqref{eq:sm_jx_full} and \eqref{eq:sm_jy_full} therefore summarize the entire symmetry-filtered transport content of the local model.

Three experimentally useful driving schemes follow immediately. (i) For a purely longitudinal drive along the untextured direction,
\begin{equation}
E=(E_x,0):\qquad j_x=\sigma_{xx}^{\mathrm D}E_x,\qquad j_y=\chi_{yxx}^{\mathrm{BCD}}E_x^2.
\end{equation}
This is the short-circuit nonlinear Hall configuration emphasized in the main text: the longitudinal channel remains purely Drude to this order, while the transverse response is entirely second order. (ii) For a drive along the textured direction,
\begin{equation}
E=(0,E_y):\qquad j_x=0,\qquad j_y=\bigl(\sigma_{yy}^{\mathrm D}+\delta\sigma_{yy}^{\mathrm{mix}}\bigr)E_y.
\end{equation}
This scheme isolates the linear longitudinal correction generated by the mixed phase-space curvature. In the minimal Hall-bar patch there is no accompanying Hall current. (iii) For an open transverse circuit under $E=(E_x,0)$, one imposes $j_y=0$ rather than fixing $E_y=0$. The sample then develops a compensating second-harmonic transverse field,
\begin{equation}
E_y^{2\omega}=-\frac{\chi_{yxx}^{\mathrm{BCD}}}{\sigma_{yy}^{\mathrm D}+\delta\sigma_{yy}^{\mathrm{mix}}}\,E_x^2.
\end{equation}
This is the local Hall-voltage form of the same nonlinear coefficient. The numerator is controlled by $mw_x$, while the denominator is dominated by the Drude conductivity with a smaller mixed-curvature renormalization proportional to $m\calB_s$.

It is useful to state the interpretation of Eqs.~\eqref{eq:sm_jx_full} and \eqref{eq:sm_jy_full} in plain language. A nonzero $\delta\sigma_{yy}^{\mathrm{mix}}$ diagnoses the phase-space geometry associated with the pseudogauge gradient $\partial_yA_x$. A nonzero $\chi_{yxx}^{\mathrm{BCD}}$ instead diagnoses a Berry-curvature dipole, which in the minimal model appears only when the local cone also acquires a valley-odd tilt. The linear and nonlinear signals can therefore vary independently across a textured device.

The full set of surviving coefficients is summarized in Table~\ref{tab:sm_summary}. For clarity we keep the explicit formulas here in terms of the local-patch parameters $(v_x,v_y,m,\mu,\calB_s,w_x)$, with $g_s=g_v=2$ for TBG.

\begin{table}[t]
\caption{Surviving transport coefficients in the minimal local Hall-bar cone.}
\label{tab:sm_summary}
\centering
\renewcommand{\arraystretch}{1.15}
\begin{tabular}{@{}lll@{}}
\toprule
Coefficient & Explicit value & Role \\
\midrule
$\sigma_{xx}^{\mathrm D}$
& $\displaystyle \frac{g_sg_v e^2\tau}{4\pi}\frac{v_x}{v_y}\frac{\mu^2-m^2}{\mu}$
& Drude conductivity along $x$ \\
$\sigma_{yy}^{\mathrm D}$
& $\displaystyle \frac{g_sg_v e^2\tau}{4\pi}\frac{v_y}{v_x}\frac{\mu^2-m^2}{\mu}$
& Drude conductivity along $y$ \\
$\delta\sigma_{yy}^{\mathrm{mix}}$
& $\displaystyle \sigma_{yy}^{\mathrm D}\frac{v_xv_y m\,\calB_s}{2\mu^3}$
& mixed-curvature correction in the textured channel \\
$\sigma_{xy},\,\sigma_{yx}$
& $0$
& forbidden by mirror symmetry and two-valley cancellation \\
$\chi_{yxx}^{\mathrm{BCD}}$
& $\displaystyle \frac{g_sg_v\,3e^3\tau mw_x}{8\pi}\frac{\mu^2-m^2}{\mu^4}$
& nonlinear Hall coefficient from the BCD \\
$\chi_{xxy}^{\mathrm{BCD}}=\chi_{xyx}^{\mathrm{BCD}}$
& $\displaystyle -\frac12\chi_{yxx}^{\mathrm{BCD}}$
& crossed quadratic response \\
$\chi_{xxx},\chi_{xyy},\chi_{yyx},\chi_{yxy},\chi_{yyy}$
& $0$
& absent dynamically in the minimal cone \\
\bottomrule
\end{tabular}
\end{table}

\section{Microscopic extraction of the local-cone susceptibilities}
\label{sec:sm_bm}

This section supplies the reference parameters and the linear texture susceptibilities $X_\alpha$ of the linear texture expansion of the main text from a heterostrained Bistritzer-MacDonald (BM) model \cite{BistritzerMacDonald2011,Bi2019}, so that the tilt, pseudogauge, and scalar scales used in the main text are computed rather than assumed.

\emph{Model.}~Each layer $l=1,2$ is deformed by $E_1=+\tfrac{\theta}{2}J+u^-$ and $E_2=-\tfrac{\theta}{2}J-u^-$, where $J$ is the $90^\circ$ rotation generator and $u^-$ is the layer-relative strain defined in the main text, shared antisymmetrically between the layers. Under $\bmr\to(1+E_l)\bmr$ the monolayer Dirac points $K_j^{(\nu)}=\nu R(2\pi j/3)\,(k_D,0)$, with $k_D=4\pi/3a_0$, map onto the dressed corner images
\begin{align}
\mathcal K_{l,j}&=\bigl(1-E_l^{\mathsf T}\bigr)K_j^{(\nu)}+\bm{\mathcal A}_l,\notag\\
\bm{\mathcal A}_l&=(-1)^{l+1}\,\nu\,\frac{\sqrt3\,\beta}{2a_0}\,\bigl(u^-_{xx}-u^-_{yy},\,-2u^-_{xy}\bigr),
\label{eq:app_corners}
\end{align}
where $\bm{\mathcal A}_l$ is the intralayer strain pseudogauge field \cite{Vozmediano2010} and $\beta=3.14$. The interlayer momentum transfers are $\bm q_j=\mathcal K_{1,j}-\mathcal K_{2,j}$, the moir\'e reciprocal lattice is spanned by $\bm g_{1,2}=\bm q_{1,2}-\bm q_0$, and the tunneling matrices are $T_j=w_{AA}\one_2+w_{AB}[\cos(2\pi j/3)\sigma_x+\nu\sin(2\pi j/3)\sigma_y]$ with $(w_{AA},w_{AB})=(88,110)$~meV, i.e. a lattice-relaxation ratio $\kappa=w_{AA}/w_{AB}=0.8$ \cite{NamKoshino2017,CarrExact2019}. Intralayer blocks are $\hbar v_F\,\bms_\nu\!\cdot\!(1+E_l^{\mathsf T})\bm\kappa_l$ with $v_F=10^6$~m/s and $\bms_\nu=(\nu\sigma_x,\sigma_y)$, and hBN alignment is modeled as a staggered potential $\Delta\sigma_z$ on layer 1. We work at $\theta_{\rm ref}=1.3^\circ$ with a four-shell plane-wave cutoff ($324\times324$ Hamiltonian). Uniaxial heterostrain of magnitude $\varepsilon$ along the direction $\varphi$ enters as $u^-=\tfrac{\varepsilon}{2}R(\varphi)\,\mathrm{diag}(1,-\nu_P)R(-\varphi)$ with Poisson ratio $\nu_P=0.16$, i.e. $(u_0,u_1,u_2)=\tfrac{\varepsilon}{4}\bigl(1-\nu_P,(1+\nu_P)\cos2\varphi,(1+\nu_P)\sin2\varphi\bigr)$.

\emph{Extraction and validation.}~For each texture configuration, the gap minimum of the two central bands near the moir\'e corner hosting the layer-1 cone is located by Newton iteration on a fitted quadratic form. On a grid of half-width $0.04\,k_\theta$ around it, with $k_\theta=2k_D\sin(\theta/2)$, the band sum and the squared band splitting are fit including cubic terms (which absorb the trigonal warping), yielding the local-cone parameters $\varepsilon_0$, $\bm w$, $m$, $v_x$, $v_y$ of the main text and the lab-frame cone position $\bm p^*$, whose deviation from the reference defines $A_i^{(\nu)}$. Susceptibilities are Richardson-extrapolated central differences. The implementation passes several checks. Hermiticity and two-valley time reversal hold to machine precision, and the $C_{3z}$ symmetry of the pristine structure holds to $10^{-12}$. Enlarging the cutoff to five shells changes the extracted parameters by less than $10^{-9}$. For $C_{3z}$-preserving perturbations the cone stays pinned to the corner and reproduces the analytic corner-image motion, $\partial\bm A/\partial\delta\theta=(0,\tfrac12)\,k_D$ and $\partial\bm A/\partial u_0=(-1,0)\,k_D$, to four digits. The renormalized velocity falls from $v^*=1.70\times10^5$~m/s at $1.30^\circ$ to $0.35\times10^5$~m/s at $1.05^\circ$, locating the flat-band condition just below $1.05^\circ$, as expected for $\kappa=0.8$.

With $\Delta=26$~meV, within the $10-30$~meV range of hBN-induced staggered potentials \cite{Jung2015}, the reference cone has the gap $2m=20$~meV used in the main text, $v_x=v_y=1.63\times10^5$~m/s (the value used in Fig.~3 of the main text), $\varepsilon_0=+2.0$~meV, $\bm w=0$, and $53\%$ layer-1 weight. The second moir\'e corner hosts the layer-2-derived cone with a smaller induced gap, $2m_2\simeq5.3$~meV. The transport formulas of the main text apply to each valley-resolved cone separately, and the measured response is their sum, dominated near its band edge by the larger-gap cone considered here.

\begin{table*}[!t]
\caption{Linear texture susceptibilities $X_\alpha=\partial X/\partial\Phi_\alpha$ of the reference cone at $\theta_{\rm ref}=1.3^\circ$, $2m=20$~meV, valley $\nu=+1$, from the heterostrained BM model (Richardson-extrapolated central differences). Position responses $\partial A_i/\partial\Phi_\alpha$ are in units of $k_D$. The $\delta\theta$ susceptibilities are per radian and the $u_\alpha$ susceptibilities per unit strain. $A_i^{(\nu)}$ and $w_i$ are valley odd while $\varepsilon_0$, $m$, $v_a$ are valley even (verified to better than $10^{-5}$). Entries marked $0^{\dagger}$ are forbidden at linear order by the $C_{3z}$ symmetry of the reference structure and vanish within the numerical accuracy. The dominant shear-channel entries carry a residual finite-difference systematic of a few percent.}
\label{tab:bm_susc}
\begin{ruledtabular}
\begin{tabular}{lcccccccc}
$\Phi_\alpha$ & $\dfrac{\partial A_x}{\partial\Phi_\alpha}\,[k_D]$ & $\dfrac{\partial A_y}{\partial\Phi_\alpha}\,[k_D]$ & $\dfrac{\partial \varepsilon_0}{\partial\Phi_\alpha}$ (eV) & $\dfrac{\partial m}{\partial\Phi_\alpha}$ (eV) & $\dfrac{\partial v_x}{\partial\Phi_\alpha}$ (m/s) & $\dfrac{\partial v_y}{\partial\Phi_\alpha}$ (m/s) & $\dfrac{\partial w_x}{\partial\Phi_\alpha}$ (m/s) & $\dfrac{\partial w_y}{\partial\Phi_\alpha}$ (m/s)\\
\colrule
$\delta\theta$ & $0^{\dagger}$ & $+0.500$ & $+0.043$ & $+0.86$ & $+3.4\times10^{7}$ & $+3.4\times10^{7}$ & $0^{\dagger}$ & $0^{\dagger}$\\
$u_0$ & $-1.000$ & $0^{\dagger}$ & $+6.00$ & $-0.008$ & $+1.4\times10^{5}$ & $+1.4\times10^{5}$ & $0^{\dagger}$ & $0^{\dagger}$\\
$u_1$ & $+4.47$ & $-0.03$ & $0^{\dagger}$ & $0^{\dagger}$ & $-1.2\times10^{6}$ & $+1.1\times10^{6}$ & $+1.5\times10^{6}$ & $-5.86\times10^{7}$\\
$u_2$ & $-0.03$ & $-6.47$ & $0^{\dagger}$ & $0^{\dagger}$ & $-6.3\times10^{7}$ & $+6.4\times10^{7}$ & $-5.85\times10^{7}$ & $-1.5\times10^{6}$\\
\end{tabular}
\end{ruledtabular}
\end{table*}

\emph{Susceptibilities.}~Table~\ref{tab:bm_susc} collects the linear susceptibilities of all local-cone parameters with respect to $\Phi_\alpha=(\delta\theta,u_0,u_1,u_2)$. Three structural statements follow. (i)~The valley parities assumed in the main text are confirmed numerically. The shift $A_i^{(\nu)}$ and the tilt reverse with $\nu$, while $\varepsilon_0$, $m$, and $v_a$ are valley even. (ii)~The position response separates into two pieces. The first is the analytic motion of the bare corner image, Eq.~\eqref{eq:app_corners}, which combines the geometric and bare-pseudogauge contributions into $(+0.30,0)\,k_D$ per $u_1$ and $(0,-2.30)\,k_D$ per $u_2$. The second is a moir\'e-hybridization dressing that vanishes for the $C_{3z}$-preserving fields, for which the cone stays pinned, and takes the exactly $C_{3z}$-covariant traceless form $\mathrm{diag}(+4.17,-4.17)\,k_D$ on $(u_1,u_2)$. The corner-image piece is the kinematic frame contribution in the sense of the covariant tetrad analysis of the main text, while the dressing is genuine projected-band data. The hybridization amplifies the bare pseudogauge coupling $\sqrt3\beta/a_0$ by a factor $\approx4$, so shear-strain gradients source $\mathcal B_s$ far more efficiently than the bare estimate suggests. (iii)~The valley-odd tilt has no counterpart in an isolated uniformly strained layer and is generated entirely by the moir\'e hybridization. Its linear response takes the $C_{3z}$-covariant form $w_x+\ii w_y=\gamma_w\,(u_1-\ii u_2)$ with $\gamma_w=(0.15-5.86\,\ii)\times10^7$~m/s. The two independent cross couplings in Table~\ref{tab:bm_susc} agree to $0.2\%$, an internal consistency check of the extraction.

\begin{figure}[H]
 \centering
    \includegraphics[width=0.6\linewidth]{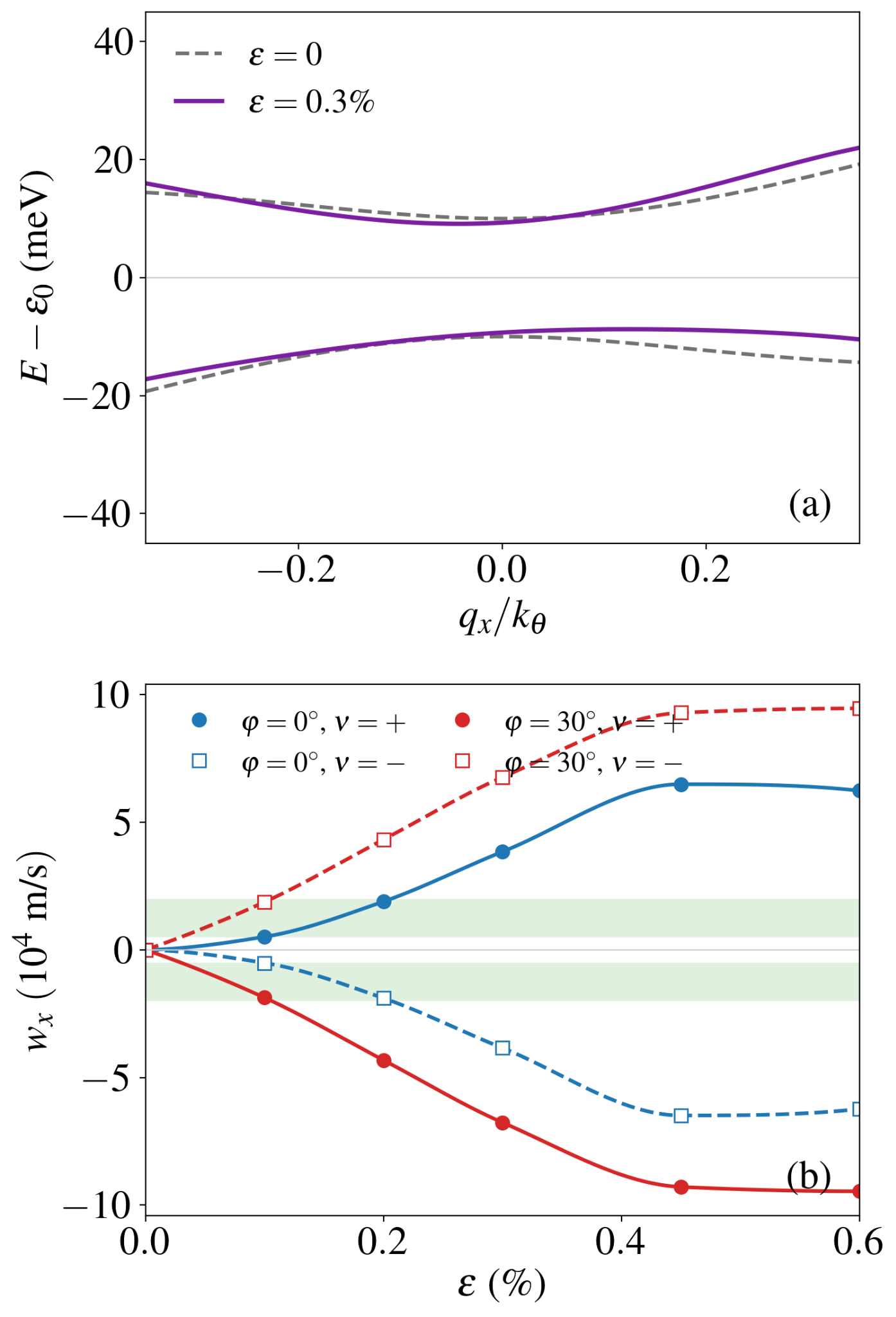}
    \caption{Microscopic anchoring of the local cone at $\theta_{\rm ref}=1.3^\circ$. (a)~Central BM bands through the gapped mini-Dirac cone along $q_x$, for the pristine reference (dashed) and for a uniaxial heterostrain $\varepsilon=0.3\%$ at $\varphi=0$ (solid, valley $\nu=+$). The strain shifts the cone, leaves the gap nearly intact ($2m=18.6$~meV), and tilts the dispersion. (b)~Extracted valley-odd tilt $w_x(\varepsilon)$ for two strain orientations and both valleys, with filled and open symbols for $\nu=\pm$. The shaded band marks the range $w_x=0.5-2\times10^4$~m/s used in the transport estimates. The exact sign reversal between the valleys confirms the valley-odd character required for the Berry-curvature dipole.}
    \label{fig4}
\end{figure}

\emph{Anchoring the transport estimates.}~For finite uniaxial heterostrain [Fig.~\ref{fig4}(b)] the tilt reaches the range assumed in the main text at sub-$0.1\%$ strain. It equals $|w_x|=(0.5,1,2)\times10^4$~m/s at $\varepsilon=(0.10,0.14,0.21)\%$ for $\varphi=0$, already at $\varepsilon=(0.03,0.06,0.11)\%$ for $\varphi=30^\circ$, and a representative $\varepsilon=0.3\%$ gives $|w_x|=3.8-6.8\times10^4$~m/s. Residual heterostrain of $0.1\%-0.7\%$ is routinely imaged in TBG devices \cite{Kerelsky2019,Huder2018,Mesple2021,Kazmierczak2021}, so the range used in Figs.~2 and 3 of the main text sits at the conservative lower end of the microscopic response. For the pseudomagnetic field, Table~\ref{tab:bm_susc} gives $\mathcal B_s=10^{14}$~m$^{-2}$ from a twist gradient $|\partial_y\theta|\simeq0.7^\circ/\mu$m (the corner-image coefficient is $\tfrac12k_D\sin\varphi_K$ for a moir\'e corner at angle $\varphi_K$ to the bar axis) or from a shear-strain gradient $\partial_yu_1\simeq1.3\times10^{-3}/\mu$m, that is $\partial_y\varepsilon\simeq0.45\%/\mu$m at $\varphi=0$. Imaged twist and strain textures vary by $0.1^\circ-0.3^\circ$ and $0.1\%-0.6\%$ over $0.1-1\,\mu$m \cite{Uri2020,Kazmierczak2021,Hu2024TwistMap}, comfortably supplying either channel. Finally, the scalar susceptibilities $\partial\varepsilon_0/\partial u_0=6.0$~eV and $\partial\varepsilon_0/\partial\delta\theta=43$~meV/rad quantify the band-edge landscape entering the local band-edge shift of the main text.

\emph{Limitations.}~The extraction is single particle. Lattice relaxation enters only through $\kappa$, interaction effects that renormalize $v^*$ and the gap are omitted, and the layer-antisymmetric deformation potential and the $O(\beta\varepsilon)$ intralayer velocity corrections are likewise neglected \cite{Vozmediano2010}. These refinements shift the numbers at the tens-of-percent level but do not affect the symmetry structure, namely the valley parities, the $C_{3z}$ selection rules, and the covariant forms above, on which the transport analysis of the main text rests.

\end{document}